\documentclass[aps,prd,amsmath,amssymb,amsfonts,floats,floatfix,superscriptaddress,nofootinbib,notitlepage]{revtex4-1}

\usepackage{hyperref}
\usepackage{graphicx}
\usepackage{graphics}
\usepackage[retainorgcmds]{IEEEtrantools}
\usepackage{color}
\usepackage{xspace} % Sensible space treatment at end of simple macros

\newcommand{\half}{\frac{1}{2}}
\newcommand{\thalf}{{\textstyle\half}}
\newcommand{\rbar}{{\bar{\mathsf{r}}}}
\newcommand{\sbar}{{\bar{\mathsf{s}}}}
\allowdisplaybreaks[1]

\begin{document}
\title{Generic effective source for scalar self-force calculations}

\author{Barry Wardell}
\affiliation{Max-Planck-Institut f\"{u}r Gravitationphysik, Albert-Einstein-Institut, 14476 Potsdam, Germany}
\affiliation{School of Mathematical Sciences and Complex \& Adaptive Systems Laboratory,\\ University College Dublin, Belfield, Dublin 4 Ireland}

\author{Ian Vega}
\affiliation{Department of Physics, University of Guelph, Guelph, Ontario, N1G 2W1, Canada}

\author{Jonathan Thornburg}
\affiliation{Department of Astronomy and Center for Spacetime Symmetries, Indiana University, Bloomington, Indiana 47405, USA}

\author{Peter Diener}
\affiliation{Center for Computation \& Technology, Louisiana State University, Baton Rouge, Louisiana 70803, USA}
\affiliation{Department of Physics \& Astronomy, Louisiana State University, Baton Rouge, Louisiana 70803, USA}

\begin{abstract}
A leading approach to the modelling of extreme mass ratio inspirals involves the treatment of the smaller mass as a point particle and the computation of a regularized self-force acting on that particle. In turn, this computation requires knowledge of the regularized retarded field generated by the particle. A direct calculation of this regularized field may be achieved by replacing the point particle with an effective source and solving directly a wave equation for the regularized field. This has the advantage that all quantities are finite and require no further regularization. In this work, we present a method for computing an effective source which is finite and continuous everywhere, and which is valid for a scalar point particle in arbitrary geodesic motion in an arbitrary background spacetime. We explain in detail various technical and practical considerations that underlie its use in several numerical self-force calculations. We consider as examples the cases of a particle in a circular orbit about Schwarzschild and Kerr black holes, and also the case of a particle following a generic time-like geodesic about a highly spinning Kerr black hole. We provide numerical \emph{C} code for computing an effective source for various orbital configurations about Schwarzschild and Kerr black holes.

\end{abstract}
\maketitle

\section{Introduction}
There has been much recent interest in the study of Extreme Mass Ratio Inspiral (EMRI) systems. These systems typically involve a compact, solar mass object inspiralling into an approximately million solar mass black hole. Such massive black holes are expected to exist at the center of most galaxies \cite{Magorrian:1997hw}.

EMRIs are expected to provide a strong source of gravitational waves for future generations of gravitational wave detectors \cite{Gair:2004,AmaroSeoane:2007,Gair:2009}.
There is also hope that parameters for these sources can be accurately estimated, enabling studies and measurements of the strong field region of central supermassive black holes \cite{Barack:Cutler:2004,Babak:Gair:Sesana:2010,Gair:Sesana:Berti:2010}. In order to achieve accurate parameter estimation, it is essential that highly accurate gravitational waveforms are available. This, in turn requires highly accurate, long-time models of the inspiral. 

A leading approach to the accurate modelling of EMRI systems arises from the fact that the mass ratio, $\mu$, is very small. This makes it possible to treat the system within perturbation theory, in which the smaller object is assumed to be a point particle generating a perturbation about the background of the larger mass. At zeroth order in $\mu$, the smaller object merely follows a geodesic of the background. At first order, it deviates from this geodesic due to its interaction with its self-field. This deviation may be viewed as a force acting on the smaller object, referred to as the \emph{self-force}. The calculation of this self-force is critical to the accurate modelling of the evolution of the system.

A na\"ive calculation of the first order perturbation leads to a retarded field which diverges at the location of the particle. The self-force, being the derivative of the field, therefore also diverges at the location of the particle and must be regularized. A series of derivations of the regularized first order equations of motion (now commonly referred to as the MiSaTaQuWa equations, named after
    Mino, Sasaki, Tanaka \cite{Mino:Sasaki:Tanaka:1996} and Quinn and Wald \cite{Quinn:Wald:1997} who first derived them) for a point particle in curved spacetime have been developed \cite{Dirac-1938,DeWitt:1960,Hobbs:1968a,Mino:Sasaki:Tanaka:1996,Quinn:Wald:1997,Quinn:2000,Detweiler-Whiting-2003,Harte-2008,Harte:2009,Harte:2010}, culminating in a recent rigorous work by Gralla and Wald \cite{Gralla:Wald:2008} and Pound \cite{pound:2009} in the gravitational case and by Gralla, et al. \cite{Gralla:Harte:Wald:2009} in the electromagnetic case. Several practical computational strategies have developed from these formal derivations:
\begin{itemize}
\item By measuring the flux of gravitational waves onto the horizon of the larger black hole and out to infinity, a time-averaged dissipative component of the self-force - which is finite and does not require regularization - may be computed. This, however, neglects potentially important conservative effects which may significantly alter the orbital phase of the system.
\item The \emph{mode-sum} approach, introduced in Refs.~\cite{Barack:Ori:2000,Barack:Mino:Nakano:Ori:Sasaki:2001}, which involves the decomposition of the retarded field into spherical harmonic modes (which are finite, but not differentiable at the particle), solving for each mode independently and subtracting ``regularization parameters'', then summing over modes. This method has been used to compute the self force for a variety of configurations in the Schwarzschild \cite{Barack:Burko:2000,Burko:2000b,Detweiler:Messaritaki:Whiting:2002,DiazRivera:2004,Haas:Poisson:2006,Haas:2007,Canizares:Sopuerta:2009,Canizares:Sopuerta:Jaramillo:2010,Barack:Sago:2007,Barack:Lousto:2002,Sago:Barack:Detweiler:2008,Detweiler:2008,Sago:2009,Barack:Sago:2010,Keidl:Shah:Friedman:Kim:Price:2010,Shah:Keidl:Friedman:Kim:Price:2010} and Kerr \cite{Warburton:Barack:2009,Warburton:Barack:2010,Thornburg:2010} spacetimes.
\item The \emph{effective source} approach \cite{vega-etal:11,diener-etal:12a,Barack:Golbourn:2007,Vega:Detweiler:2008,Vega:2009,Dolan:Barack:2010} in which the regularization is done before solving the wave equation. In this case, all quantities are finite throughout the calculation and one directly solves a wave equation for the regularized field. A review of this approach can be found in \cite{vega-etal:11}. Note that the effective source proposed by Lousto and Nakano \cite{Lousto:Nakano:2008} differs in that it is not derived from the Detweiler-Whiting singular field.
\item The \emph{matched expansion} approach \cite{Anderson:Wiseman:2005,Casals:Dolan:Ottewill:Wardell:2009} in which a quasi-local expansion of the Green function \cite{Ottewill:Wardell:2008,Ottewill:Wardell:2009,QL} (which is valid in the recent past) is matched onto a quasi-normal mode sum (valid in the distant past)\footnote{In black hole spacetimes there is also a branch cut integral which must be evaluated in the region where the quasi-normal mode sum is used. Substantial recent progress has been made towards the calculation of this branch cut contribution \cite{Casals:2011aa}.}. The retarded field is then computed as the integral of this matched retarded Green function along the worldline of the particle.
\end{itemize}
For a comprehensive review of the self-force problem, see Refs.~\cite{Poisson:2003,Detweiler:2005,Barack:2009}. The present work focuses on the third of these strategies, the \emph{effective source} approach. In this approach, the point particle source is replaced with a finite effective source leading to a wave equation which admits the correct regularized field (at the particle) as a solution.

Given our motivation in studying the EMRI problem, it is the gravitational self-force which is of the most interest. In this paper, however, we instead study the analogous \emph{scalar} self-force. This allows us to develop insight and techniques without being obscured by the additional complexity of the gravitational case. It should be noted, however, that this extra complexity is predominantly only calculational and comes in the form of larger expressions. Conceptually, the calculations done here follow through for the gravitational case with few modifications.

The purpose of this paper is to provide a comprehensive exposition of the scalar effective source employed in a variety of 
recent and ongoing self-force calculations \cite{Dolan:Barack:2010,dolan-etal:11,diener-etal:12a,diener-etal:12b}. 
Starting with its covariant definition, we present 
its coordinate construction and the various modifications that we found necessary in order to get the effective 
source to its current ``best" form. Much of the paper is technical in nature, but we believe that all the details 
provided here are essential to anyone interested in pursuing an effective source approach to self-force calculations. 

The layout of the paper is as follows. In Sec.~\ref{sec:effsource} we introduce the effective source approach in detail and compute approximations to the singular field and effective source in the form of covariant expansions. In Sec.~\ref{sec:practical}, we develop practical methods for evaluating these approximations in a specific spacetime in terms of coordinate expansions. We give example calculations for the case of a circular geodesic orbit in Schwarzschild and Kerr spacetimes and a generic orbit in Kerr spacetime in Sec.~\ref{sec:examples}. In Sec.~\ref{sec:discussion} we conclude with a discussion on aspects of the calculation and on prospects for future applications. In Appendix \ref{sec:Expansions}, we develop covariant expansions of various biscalars used in this paper. In Appendix \ref{sec:s2positive}, we discuss a modification to the covariant expansion which yields substantial practical benefits. Finally, in Appendix \ref{sec:numerical}, we discuss issues related to efficient numerical implementations for computing the singular field and effective source.

Many of the expressions developed in this work, although useful, are too unwieldy to be given in printed form. Instead, we have made available all expressions we deem to be useful online \cite{EffSource-online} as \emph{Mathematica} code. Furthermore, as this work is intended to provide computational tools for those interested in doing self-force calculations, we also include a library of \emph{C} code for computing the singular field and effective source for various configurations in Schwarzschild and Kerr spacetimes. The intention is for this code to be a ``black box'' which can be easily incorporated into existing numerical codes, whether they are $3+1$D, $2+1$D or $1+1$D.

Throughout this paper, we use units in which $G=c=1$ and adopt the sign conventions of \cite{Misner:Thorne:Wheeler:1974}. We denote symmetrization of indices using brackets (e.g. $(\alpha \beta)$) and exclude indices from symmetrization by surrounding them by vertical bars (e.g. $(\alpha | \beta | \gamma)$). Roman letters are used for free indices and Greek letters for indices summed over all spacetime dimensions. Roman letters starting from $i$ are used for indices summed only over spatial dimensions. Capital letters are used to denote the spinorial/tensorial indices appropriate to the field being considered. For convenience, we frequently make use of the shorthand notation of Ref.~\cite{Haas:Poisson:2006} by introducing definitions such as $R_{u \sigma u \sigma | \sigma} \equiv R_{\bar{\alpha} \bar{\beta} \bar{\gamma} \bar{\delta} ; \bar{\epsilon}} u^{\bar{\alpha}} \sigma^{\bar{\beta}} u^{\bar{\gamma}} \sigma^{\bar{\epsilon}} \sigma^{\bar{\delta}}$.

\section{Effective source approach}
\label{sec:effsource}
To compute the self-force, $f^\alpha$, acting on a point particle with scalar charge $q$, knowledge of the retarded field, $\Phi_{\rm ret}$, generated by the particle is required. This field is a solution of the inhomogeneous wave equation,
\begin{equation}
\label{eq:wave}
\mathcal{D} \Phi_{\rm ret}(x) = -4 \pi q \int_\gamma \frac{\delta^4(x-z(\tau))}{\sqrt{-g}} d\tau,
\end{equation}
where the source corresponds to a point particle on a worldline $\gamma$ in some background spacetime and where
\begin{equation}
\label{eq:wave-operator}
\mathcal{D} \equiv (\Box - \xi R)
\end{equation}
is the scalar wave operator. A na\"ive calculation of the self-force from this retarded field will diverge when evaluated at the location of the particle. In order to compute a meaningful self-force, one must therefore find a regularized retarded field. This may be achieved by separating the field into singular (S) and regular (R) parts,
\begin{equation}
\label{eq:split-field}
\Phi_{\rm ret} = \Phi_{\rm S} + \Phi_{\rm R}.
\end{equation}
The identification of a singular field which gives the correct regularized self-force is crucial. Using a Green function decomposition, Detweiler and Whiting \cite{Detweiler-Whiting-2003} were able to find a representation of the singular field which is valid in a region near to the particle. It is a solution of the same inhomogeneous wave equation \eqref{eq:wave} as the retarded field. A brief overview of their approach is given in the next subsection.

Given knowledge of the singular field, one must then prescribe a method of computing the regularized field. In the effective source approach, first proposed independently by Barack and Golbourn \cite{Barack:Golbourn:2007} and by Vega and Detweiler \cite{Vega:Detweiler:2008}, the splitting of the self-field into regular and singular parts is done at the level of the wave equation, 
\begin{equation}
\mathcal{D} \Phi_{\rm ret} = \mathcal{D} \Phi_{\rm S} + \mathcal{D} \Phi_{\rm R},
\end{equation}
\emph{before} solving for the field. One then solves directly the equation for the regularized field.
\begin{equation}
\mathcal{D} \Phi_{\rm R} = -4 \pi q \int_\gamma \frac{\delta^4(x-z(\tau))}{\sqrt{-g}} d\tau - \mathcal{D} \Phi_{\rm S} = 0.
\end{equation}
The regularized self-force is then simply given by the derivative of this regularized field,
\begin{equation}
f^a = q \nabla^a \Phi_{\rm R}.
\end{equation}
This method has several advantages:
\begin{itemize}
\item It does not rely on the separability of the field equations. This is particularly important in the Kerr spacetime where the perturbation equations are not fully separable in the time domain.  
\item There are no troublesome delta functions or singularities to deal with. This is particularly advantageous in numerical calculations where smoothness is desirable.
\item In comparison to methods which first compute $\Phi_{\rm ret}$ and then regularize, there is no need to cancel two large quantities ($\Phi_{\rm S}$ and $\Phi_{\rm ret}$) to get the self-force, so in principle the field one solves for is inherently more accurate. This is particularly relevant in numerical calculations, where the cancellation of large quantities may lead to considerable round-off errors.
\item Its applicability in the time domain means that the orbit may be evolved, coupling the geodesic equations into the wave equation and source calculation.
\end{itemize}

In principle, the regularized field is a solution of the homogeneous wave equation and the self-force is determined purely from the boundary condition for $\Phi_{\rm R} = \Phi_{\rm ret} - \Phi_{\rm S}$. However, in practice the singular field identified by Detweiler and Whiting is not defined globally (it is not even clear that a global definition exists). One must therefore introduce a method for restricting the singular field to a region near the particle. Furthermore, in practice an exact calculation of the singular field away from the particle proves difficult; it is much easier to calculate an approximation to the singular field, denoted by $\tilde{\Phi}_{\rm S}$ and to solve for an approximate regularized field $\tilde{\Phi}_{\rm R}$. The construction of an approximate singular field must ensure that its local expansion near the particle matches that of the actual singular field sufficiently well that evaluating the self-force using $\tilde{\Phi}_{\rm R}$ yields the correct value \emph{at} the particle. It is important to note, however, that this approximate regularized field becomes meaningless far from the particle.

There are two different approaches to dealing with the problem of the lack of a global definition for the singular field. Vega and Detweiler \cite{Vega:Detweiler:2008} tackle the issue of restricting the singular field to a region near the particle with the use of a \emph{window function}, $W$, and split the retarded field as
\begin{equation}
\Phi_{\rm ret} = W \tilde{\Phi}_{\rm S} + \tilde{\Phi}_{\rm R}.
\end{equation}
The window function is chosen so that near to the particle $W \tilde{\Phi}_{\rm S}$ remains a good approximation to the singular field, while far away from the particle $W$ dies away sufficiently quickly that $\tilde{\Phi}_{\rm R} \approx \Phi_{\rm ret}$. Barack and Golbourn \cite{Barack:Golbourn:2007} take an alternative approach. They introduce a world-tube around the particle. Inside the world-tube, they solve for $\tilde{\Phi}_{\rm R}$ and outside they solve for $\Phi_{\rm ret}$. They then impose \eqref{eq:split-field} as what is essentially a ``change of variables'' on the world-tube boundary\footnote{In a numerical implementation, it is common to reduce the wave equation to a system of first order equations. In this case, it may be necessary to impose the conditions not only on $\tilde{\Phi}_{\rm R}$, but also on its derivatives.}. In this case, the lack of a global definition for $\Phi_{\rm S}$ is no longer an issue as the only requirement on $\tilde{\Phi}_{\rm S}$ is that it approximates the singular field sufficiently well near the particle.

In both cases, the approximation to the singular field, $\tilde{\Phi}_{\rm S}$, is no longer a solution of Eq.~\eqref{eq:wave}. The source term now has additional structure \emph{away} from the particle, extending throughout the worldtube or the region of support of the window function.
As a result, the approximate regularized field is now a solution of the inhomogeneous wave equation with an effective source, $S_{\rm eff}$:
\begin{equation}
\mathcal{D} \tilde{\Phi}_{\rm R} = -4 \pi q \int_\gamma \frac{\delta^4(x-z(\tau))}{\sqrt{-g}} d\tau - \mathcal{D} \left(W\tilde{\Phi}_{\rm S}\right) \equiv S_{\rm eff}.
\end{equation}
In contrast to Eq.~\eqref{eq:wave}, however, this source has the advantage of being regular and smooth everywhere except at the location of the particle where it is still regular, but of finite differentiability, the level of differentiability being determined by the choice of approximation to the singular field.

\subsection{Exact expression for the singular field}
In order to obtain an expression for the singular field, we follow Detweiler and Whiting \cite{Detweiler-Whiting-2003} in introducing the Hadamard form \cite{Hadamard,Friedlander} for the singular Green function,
\begin{equation}
\label{eq:GF-S}
G_{\rm S}(x,x') = \frac{1}{2}\left[U(x,x') \delta(\sigma(x,x')) + V(x,x') \theta(\sigma(x,x'))\right],
\end{equation}
which is obtained by adding a homogeneous solution (in this case $V(x,x')$) of the wave equation to the symmetric Green function, $G_{\rm sym} = \frac{1}{2} \left( G_{\rm ret} + G_{\rm adv} \right)$ \cite{DeWitt:1965}. Here, $\delta\left( \sigma\left(x,x'\right)\right)$ is the covariant form of the Dirac delta function, $\theta\left( \sigma\left(x,x'\right)\right)$ is the Heaviside step function, and $U\left( x,x' \right)$ and $V\left( x,x' \right)$ are symmetric biscalars which are regular for $x' \rightarrow x$. The biscalar $\sigma \left( x,x' \right)$ is the Synge~\cite{Synge,Poisson:2003} world function, which is equal to one half of the squared geodesic distance between $x$ and $x'$.

\begin{figure}
\includegraphics[scale=0.5]{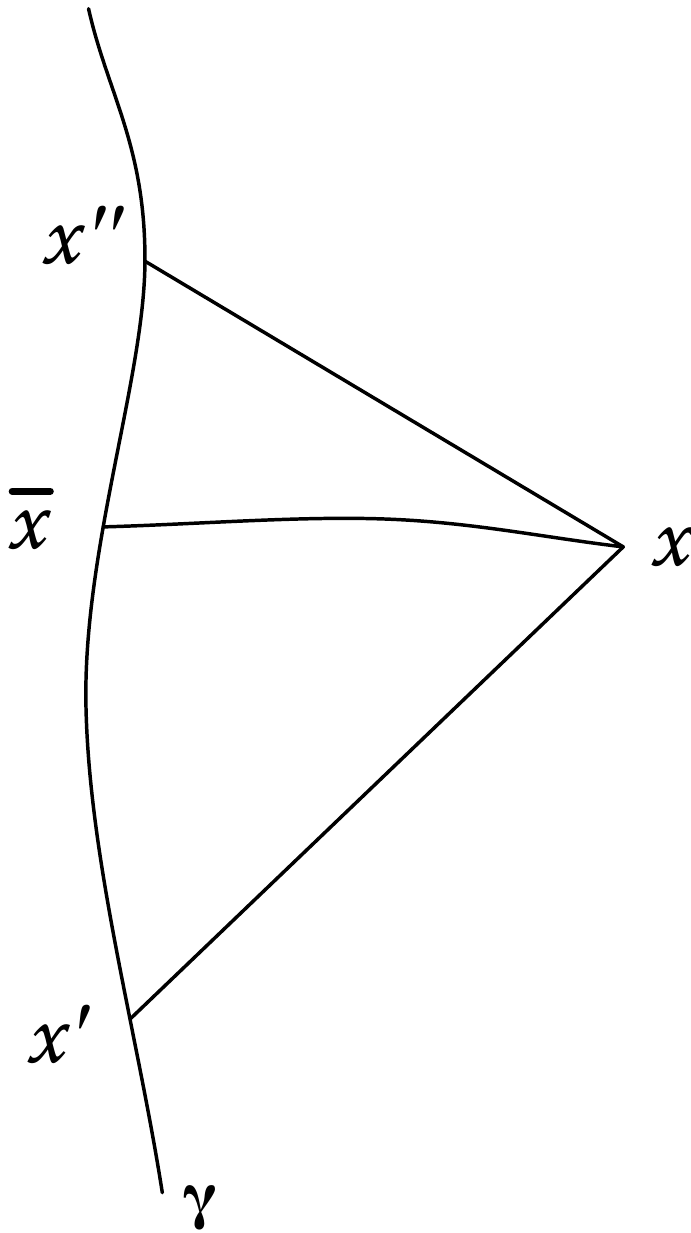}
\caption{The singular field at the point $x$ can be expressed in terms of the retarded and advanced distances to the world-line, $\gamma$.}
\label{fig:SingularField}
\end{figure}

This singular Green function is a solution of the same wave equation as the symmetric Green function, but differs in that it has support only on and \emph{outside} the light-cone. Note that this singular Green function is not guaranteed to exist globally. Its definition depends on the existence of the unique function $V(x,x')$, which is only true provided $x$ and $x'$ are within a convex normal neighborhood\footnote{When considering Hadamard form Green functions such as \eqref{eq:GF-S}, one typically defines them within a \emph{causal domain} \cite{Friedlander}. The singular Green function is acausal so this must be relaxed to a definition within a convex normal neighborhood, requiring only that $\sigma(x,x')$ be unique.}. Fortunately, in the effective source approach we only require that it exists in a neighborhood of the particle, in which case it can be given the clear definition \eqref{eq:GF-S}. We now define the singular field by
\begin{equation}
\label{eq:PhiS-integral}
\Phi_{\rm S} = \int_\gamma G_{\rm S} (x,z(\tau)) d\tau.
\end{equation}
Substituting \eqref{eq:GF-S} into \eqref{eq:PhiS-integral} and making the change of variables $\tau \rightarrow \sigma(x,z(\tau))$, we obtain an expression for the singular field which depends on a
finite portion of the particle's world line:
\begin{equation}
\label{eq:PhiS}
\Phi_{\rm S}(x) = \frac{U(x,x')}{2\sigma_{\alpha'} u^{\alpha'}} - \frac{U(x,x'')}{2\sigma_{\alpha''} u^{\alpha''}} + \frac{1}{2}\int_{u}^{v} V(x,z(\tau)) d\tau.
\end{equation}
Here we have introduced the retarded and advanced points $x'$ and $x''$ corresponding to the retarded and advanced times $u$ and $v$ on the world-line $\gamma$ associated with the field point $x$ (Fig.~\ref{fig:SingularField}).

\subsection{Approximation to the singular field}
The expression for the singular field given in Eq.~\eqref{eq:PhiS} is very general. It is valid for any worldline in any spacetime provided the field point $x$ is sufficiently close to the worldline that the singular Green function can be defined. In practice, it is only in very simple spacetimes 
 that $U(x,z(\tau))$, $\sigma(x,z(\tau))$ and $V(x,z(\tau))$ may be computed exactly. In many curved spacetimes of interest (including Schwarzschild and Kerr) this is not the case and one must find an approximation to \eqref{eq:PhiS}.

In the present work, we choose a covariant series expansion of \eqref{eq:PhiS} (taken to second order in the geodesic distance from the field point to the world line) as a starting point for our approximation to $\Phi_{\rm S}$. In doing so, we use the methods described in Refs.~\cite{Poisson:2003} and \cite{Haas:Poisson:2006} to consolidate the dependence of $\Phi_{\rm S}$ on the advanced and retarded points $x'$ and $x''$ into a single arbitrary point $\bar{x}$ on the worldline. This has the additional advantage of making the dependence of $x'(x)$ and $x''(x)$ on $x$ explicit, so that $\bar{x}$ is truly an arbitrary point on the worldline (sufficiently close to $x'$ and $x''$) with no implicit dependence on $x$. We additionally make use of the techniques of Ref.~\cite{Ottewill:Wardell:2009} to compute covariant expansions of all required bitensors.

Given the primary motivation of studying black hole spacetimes such as Schwarzschild and Kerr, it
is reasonable to assume that the spacetime is vacuum (i.e.\  $R_{ab} = 0$). However, in the present
work, we do \emph{not} make that assumption. This is motivated by the fact that some of the
leading-order terms in the covariant local expansion of the gravitational singular field involve
the Riemann tensor and do not vanish in vacuum. In contrast, the analogous leading-order terms
for the scalar case involve only the Ricci tensor\footnote{More specifically, in the scalar case
the first four orders in the covariant expansion of the tail term, $V(x,x')$, involve only the Ricci tensor
and Ricci scalar. As a result, $V(x,x')=\mathcal{O}(\epsilon^4)$ in vacuum and the tail term could
be neglected in the present calculation. In the gravitational
case, however, the tail term, $V_{aa'bb'}(x,x')$, has a leading order component involving the Riemann
tensor. This means that even in vacuum $V_{aa'bb'}(x,x')=\mathcal{O}(1)$.}. As far as the covariant local expansions are
concerned then, the scalar singular field in a \emph{non-vacuum} spacetime best captures the
structure of the gravitational singular field in a generic spacetime.
%This is motivated by the fact that in the gravitational case, even in vacuum, an approximation to the singular gravitational field will involve terms arising from the gravitational equivalents of $U(x,x')$ and $V(x,x')$ 
By not assuming vacuum, we are therefore emulating some of the extra complexity which would otherwise only appear in the gravitational case.

The covariant expansion of $\Phi_{\rm S}$ requires, in turn, the expansion of the functions $U(x,x')$, $U(x,x'')$, $\sigma_{\alpha'} u^{\alpha'}$, $\sigma_{\alpha''} u^{\alpha''}$ and $V(x,z(\tau))$ about the point $\bar{x}$. We compute these expansions in Appendix~\ref{sec:Expansions}. Substituting \eqref{eq:Upm}, \eqref{eq:r_ret}, \eqref{eq:r_adv} and \eqref{eq:intV} into \eqref{eq:PhiS}, we get
\begin{IEEEeqnarray}{rCl}
\label{eq:PhiS-approx}
\Phi_{\rm S} \approx & \tilde{\Phi}_{\rm S} &= q\Bigg\{\frac{1}{\sbar} + \bigg[\frac{\rbar^2-\sbar^2}{6\,\sbar^3} R_{u \sigma u \sigma}  + \frac{1}{12\,\sbar} \big(2\, \rbar\, R_{u\sigma} + R_{\sigma \sigma} + R_{uu}(\rbar^2+\sbar^2)\big) + \frac{1}{2}\left(\xi - \frac{1}{6}\right) \bar{R}\, \sbar \bigg] \nonumber \\
&& +\: \bigg[ \frac{1}{24\,\sbar}\big( -R_{\sigma \sigma | \sigma} + ( R_{\sigma \sigma | u} -2 R_{u \sigma | \sigma} )\,\rbar +(2R_{u \sigma | u}-R_{uu|\sigma})(\rbar^2+\sbar^2)+ R_{uu|u}(\rbar^3+3\,\rbar\,\sbar^2)\big) \nonumber \\
&& + \frac{1}{4} (\xi-\frac{1}{6})(\bar{R}_{|u} \,\rbar\,\sbar-\bar{R}_{\sigma}\,\sbar) +\frac{1}{24 \,\sbar^3} \big( \left(\rbar^2 - 3 \,\sbar^2\right) \rbar\, R_{u \sigma u \sigma | u} - \left(\rbar^2-\sbar^2\right) R_{u \sigma u \sigma | \sigma} \big) \bigg]\Bigg\}.
\end{IEEEeqnarray}
where $\sbar \equiv (g^{\bar{\alpha} \bar{\beta}} + u^{\bar{\alpha}} u^{\bar{\beta}}) \sigma_{\bar{\alpha}} \sigma_{\bar{\beta}}$  (i.e.\ the projection of $\sigma_{\bar{a}}$ orthogonal to the worldline), and $\rbar = \sigma_{\bar{\alpha}} u^{\bar{\alpha}}$ (the projection along the worldline) and we adopt the notation of Haas and Poisson \cite{Haas:Poisson:2006} in defining $R_{u \sigma u \sigma | \sigma} \equiv R_{\bar{\alpha} \bar{\beta} \bar{\gamma} \bar{\delta} ; \bar{\epsilon}} u^{\bar{\alpha}} \sigma^{\bar{\beta}} u^{\bar{\gamma}} \sigma^{\bar{\epsilon}} \sigma^{\bar{\delta}}$. Letting $\epsilon$ be a measure of the geodesic distance from $x$ to the world-line (i.e.\ $\bar{x}$), the first term here is $\mathcal{O}(\epsilon^{-1})$, the second group of terms is $\mathcal{O}(\epsilon^1)$ and the third group of terms is $\mathcal{O}(\epsilon^2)$. The difference between $\Phi_{\rm S}$ and $\tilde{\Phi}_{\rm S}$ is then $\mathcal{O}(\epsilon^3)$

\subsection{Approximation to the effective source} \label{sec:effsrc-approx}
Given the approximation \eqref{eq:PhiS-approx} to the singular field, a corresponding effective source may be computed by applying the wave operator to $\tilde{\Phi}_{\rm S}$. (This requires cancelling the divergent terms in $\tilde{\Phi}_S$, so the derivatives in the wave operator must be
computed very accurately. In particular, straightforward numerical differentiation does not provide sufficient accuracy close to the particle.) In this section, we give an exact expression for the effective source and compute an approximation which is valid near the particle. This approximation gives insight into the properties of a source derived from a particular order approximation to the singular field.

Before proceeding further, we will clarify the meaning of `order' as used in this context. All approximations are considered as expansions in powers of $\epsilon$, which is roughly speaking the distance between $x$ and the world-line (i.e.\ the length of the bivector $\sigma^{\bar{a}}$). This means that $\sbar$, $\rbar$, and $\sigma^{\bar{a}}$ are all of order $\epsilon$. The order of an approximation is then defined in terms of the order of the approximation to the singular field. The first order approximation is given by the leading term (of order $\epsilon^{-1}$) in the approximation to $\Phi_{\rm S}$, i.e.\ the first term in \eqref{eq:PhiS-approx}. The second order approximation is given by the first two orders (to order $\epsilon^0$) in the expansion of $\Phi_{\rm S}$. As there is no term at order $\epsilon^0$ in \eqref{eq:PhiS-approx}, at this stage the second order approximation is equivalent to the first order approximation. As will be discussed in Sec.~\ref{sec:practical} this will not, however, always be the case. Likewise, the third order approximation includes terms up to order $\epsilon$ in $\Phi_{\rm S}$ and the fourth order includes terms up to order $\epsilon^2$. When referring to the effective source, the order referred to will be determined by the order of the singular field from which it is derived so that the first order effective source will be given by the wave operator acting on the first order singular field, and so on.

\subsubsection{First order}
The first order approximation to the singular field is given by the leading term in \eqref{eq:PhiS-approx}:
\begin{equation}
\label{eq:PhiS-O1}
\tilde{\Phi}_{\rm S}^{(1)} = \frac{1}{\sbar},
\end{equation}
which is of order $\epsilon^{-1}$. Since the wave operator contains second derivatives, one would in general expect that the result would be of order $\epsilon^{-3}$. Applying \eqref{eq:wave-operator} to \eqref{eq:PhiS-O1} we find that this does appear to be the case:
\begin{IEEEeqnarray}{rCr}
S_{\rm eff}^{(1)} = \mathcal{D} (\frac{1}{\sbar})
= \frac{3(\rbar^2+\sbar^2)}{\sbar^5} + \frac{3\,\rbar^2-\sbar^2}{\sbar^5}\nabla_\alpha \rbar \,\nabla^\alpha \rbar - \frac{{\sigma_\alpha}^\alpha}{\sbar^3} - \frac{\rbar\,\Box\,\rbar}{\sbar^3} - \frac{\xi R}{\sbar}.
\end{IEEEeqnarray}
Noting that $\nabla_a \rbar$ and ${\sigma_\alpha}^\alpha$ are $\mathcal{O}(1)$ and $\Box\,\rbar$ is $\mathcal{O}(\epsilon)$, we see that the first three terms here are $\mathcal{O}(\epsilon^{-3})$ and the last two are  $\mathcal{O}(\epsilon^{-1})$; it would appear that the first order effective source is $\mathcal{O}(\epsilon^{-3})$. However, expanding $\nabla_a \rbar$, ${\sigma_\alpha}^\alpha$, $\Box \,\rbar$ and $R$ about $\bar{x}$, we find that the $\mathcal{O}(\epsilon^{-3})$ terms cancel, leaving a source which is $\mathcal{O}(\epsilon^{-1})$:
\begin{IEEEeqnarray}{rCl} 
S_{\rm eff}^{(1)}&=& \bigg[\frac{3\,\rbar^2-\sbar^2}{3\,\sbar^5}R_{u\sigma u \sigma} + \frac{1}{3\,\sbar^3}\left(2\,\rbar R_{u \sigma}+R_{\sigma \sigma}\right) -\frac{\xi \bar{R}}{\sbar} \bigg]\nonumber \\
&& -\: \bigg[ \frac{3\,\rbar^2 - s^2}{6\,\sbar^5}R_{u\sigma u\sigma | \sigma} - \frac{1}{12\,\sbar^3}\left(\rbar\,R_{\sigma \sigma | u} -6\,\rbar\,R_{u\sigma | \sigma} - 3R_{\sigma \sigma | \sigma}\right)- \frac{\xi \bar{R}_{|\sigma}}{\sbar} \bigg] + \mathcal{O}(\epsilon).
\end{IEEEeqnarray}
Here, the first group of terms are $\mathcal{O}(\epsilon^{-1})$ and the second group are $\mathcal{O}(1)$. In other words, the first order source diverges at the particle like $1/\epsilon$, i.e.\ it is $\mathcal{C}^{-2}$. This is sufficient to give a finite, but discontinuous regularized field. As a result of the discontinuity of the field at the particle, it is not possible to compute the self-force from its derivative.

Note that the second order source will be the same as the first order source and will therefore have the same properties, with one caveat: by decomposing into $m$-modes, Barack et al. \cite{Barack:Golbourn:Sago:2007} were able to extract a self-force from a second order source. This may be understood as a result of the ``averaging'' effect the Fourier transform used in the $m$-mode decomposition has on the smoothness of the source.

\subsubsection{Third order}
The third order approximation to the singular field is given by the two leading terms in \eqref{eq:PhiS-approx}:
\begin{equation}
\label{eq:PhiS-O3}
\Phi_{\rm S}^{(3)} = \Phi_{\rm S}^{(1)} + \bigg[\frac{\rbar^2-\sbar^2}{6\,\sbar^3} R_{u \sigma u \sigma}  + \frac{1}{12\,\sbar} \big(2 \,\rbar\, R_{u\sigma} + R_{\sigma \sigma} + R_{uu}(\rbar^2+\sbar^2)\big) + \frac{1}{2} \left(\xi - \frac{1}{6}\right) \bar{R}\, \sbar \bigg].
\end{equation}
Applying \eqref{eq:wave-operator} gives the third order effective source:
\begin{IEEEeqnarray}{rCl}
S_{\rm eff}^{(3)} &=& S_{\rm eff}^{(1)} + \frac{1}{12\,\sbar^7}\biggl\{\rbar \,\sbar^{4} \Bigl[ \sbar^{2}(\rbar \, \Box \,\rbar + \Box \sigma) - (\rbar^{2} +  \sbar^{2}) -  (\rbar^{2} - \sbar^{2})  \nabla_\alpha \rbar\, \nabla^\alpha \rbar\Bigr] \Bigl[-1 + 6 \xi \Bigr] \nonumber \\ 
&& - \: R{}_{\sigma }{}_{\sigma } \sbar^{2} \Bigl[\rbar \,\sbar^{2}  \Box\, \rbar + \sbar^{2}  \Box\sigma -  (1 + \nabla_\alpha \rbar\, \nabla^\alpha \rbar) \bigl(3\, \rbar^{2} -  \sbar^{2}\bigr)\Bigr] + 2 R{}_{\sigma }{}_{\bar{\alpha}} \sbar^4 \Bigl[ \sbar^{2} \sigma {}^{\bar{\alpha}}{}^{\beta}{}_{\beta} - 2 \,\rbar\, \sigma {}^{\bar{\alpha}}{}_{\beta} \nabla^\beta \rbar\Bigr] \nonumber \\ 
&&   + 2 R{}_{u}{}_{\sigma }\sbar^{2} \Bigl(\sbar^2 \bigl(\sbar^2 - \rbar^{2}\bigr) \Box\,\rbar  + \rbar \bigl(3 \,\rbar^{2} - \sbar^{2} - \sbar^{2} \Box \sigma + 3 (\rbar^{2} - \sbar^{2})\bigr)\nabla_\alpha \rbar\, \nabla^\alpha \rbar \Bigr)\nonumber \\ 
&&   + R{}_{u}{}_{u}\sbar^{2} \Bigl[\rbar\,\sbar^2 \bigl(3\sbar^{2}- \rbar^{2} \bigr)\Box\,\rbar + \bigl(\rbar^{2} - \sbar^{2}\bigr)\bigl(3 \,\rbar^{2} +\sbar^{2} - \sbar^{2} \Box \sigma + 3 \bigl(\rbar^{2} - \sbar^{2}\bigr) \nabla_\alpha \rbar\, \nabla^\alpha \rbar \Bigr)\Bigr] \nonumber \\ 
&& +\: 2\,\sbar^4\, R{}_{u}{}_{\bar{\alpha}} \Bigl[ \rbar\,\sbar^{2} \sigma {}^{\bar{\alpha}}{}^{\beta}{}_{\beta} + 2 \bigl(\sbar^{2}- \rbar^{2}\bigr) \sigma {}^{\bar{\alpha}}{}_{\beta} \nabla^\beta \rbar \Bigr]  + 2 R{}_{\bar{\alpha} \bar{\beta}}\,\sbar^{6} \sigma {}^{\bar{\beta}}{}_{\gamma} \sigma {}^{\bar{\alpha}}{}^{\gamma}\nonumber \\ 
&& + 2 R{}_{u}{}_{\sigma }{}_{u}{}_{\sigma } \Bigl[15 \rbar^{4} - 12 \rbar^{2}\,\sbar^{2} +\sbar^{4} -3  \rbar\,\sbar^2 \bigl( \rbar^{2} -\sbar^{2}\bigr) \Box\,\rbar \nonumber \\
&& \qquad -\sbar^2 \bigl( 3 \,\rbar^{2} -\sbar^{2}\bigr) \Box \sigma  + 3 \bigl( 5\, \rbar^{4} - 6 \,\rbar^{2}\,\sbar^{2} +\sbar^{4}\bigr) \nabla_\alpha \rbar \,\nabla^\alpha \rbar\Bigr]\nonumber \\ 
&& +\: 4 R{}_{u}{}_{\sigma }{}_{u}{}_{\bar{\alpha}}\sbar^{2} \Bigl[\sbar^{2} \sigma {}^{\bar{\alpha}}{}^{\beta}{}_{\beta} - 6\, \rbar\, \sigma {}^{\bar{\alpha}}{}_{\beta} \nabla^\beta \rbar\Bigr] \Bigl[ \rbar^{2} -\sbar^{2}\Bigr] + 4 R{}_{u}{}_{\bar{\alpha}}{}_{u}{}_{\bar{\beta}}\,\sbar^{4}\, \sigma {}^{\bar{\alpha}}{}_{\gamma} \sigma {}^{\bar{\beta}}{}^{\gamma} \Bigl[\rbar^{2} - \sbar^{2}\Bigr]\biggr\},
\end{IEEEeqnarray}
which appears to have an additional contribution at $\mathcal{O}(\epsilon^{-1})$ compared to the first order case. However, as before, re-expanding higher derivatives of $\sigma$ about $\bar{x}$, we find that this exactly cancels the $\mathcal{O}(\epsilon^{-1})$ contribution from the first order source, leaving a source which has a directional dependence at $\mathcal{O}(1)$:
\begin{IEEEeqnarray}{rCl}
S_{\rm eff}^{(3)} &=& S_{\rm eff}^{(1)} - \bigg[\frac{3\,\rbar^2-\sbar^2}{3\,\sbar^5}R_{u\sigma u \sigma} + \frac{1}{3\,\sbar^3}\left(2\,\rbar R_{u \sigma}+R_{\sigma \sigma}\right) -\frac{\xi \bar{R}}{\sbar} \bigg] + \mathcal{O}(\epsilon)\nonumber \\
&=& -\: \bigg[ \frac{3\,\rbar^2 - \sbar^2}{6\sbar^5}R_{u\sigma u\sigma | \sigma} - \frac{1}{12\sbar^3}\left(\rbar\,R_{\sigma \sigma | u} -6\,\rbar\,R_{u\sigma | \sigma} - 3R_{\sigma \sigma | \sigma}\right)- \frac{\xi \bar{R}_{|\sigma}}{\sbar} \bigg] + \mathcal{O}(\epsilon).
\end{IEEEeqnarray}
In this way, we see that including the third order contribution to the singular field gives a source which is $\mathcal{C}^{-1}$. This is now sufficient to calculate both the regularized field and its derivative (i.e.\ the self-force).

\subsubsection{Fourth order}
Following the procedure once more, by including the fourth order contribution to $\Phi_{\rm S}$, computing the associated effective source and re-expanding higher derivatives of $\sigma$ about $\bar{x}$, we find that it has a contribution at $\mathcal{O}(1)$ which exactly cancels that from the third order source, leaving a source which has a directional dependence at $\mathcal{O}(\epsilon)$:
\begin{IEEEeqnarray}{rCl}
S_{\rm eff}^{(4)} = S_{\rm eff}^{(3)} + \bigg[ \frac{3\,\rbar^2 - \sbar^2}{6\,\sbar^5}R_{u\sigma u\sigma | \sigma} - \frac{1}{12\,\sbar^3}\left(\rbar\,R_{\sigma \sigma | u} -6\,\rbar\,R_{u\sigma | \sigma} - 3R_{\sigma \sigma | \sigma}\right)- \frac{\xi \bar{R}_{|\sigma}}{\sbar} \bigg] + \mathcal{O}(\epsilon) =  \mathcal{O}(\epsilon) .
\end{IEEEeqnarray}
Therefore, including the fourth order contribution to the singular field we obtain a source which is $\mathcal{C}^{0}$. This is not only sufficient to give the self-force, but gives reasonably good convergence in numerical calculations.

\subsubsection{Higher orders}
\label{sec:no-effsource}
\begin{figure}
\includegraphics[width=8cm]{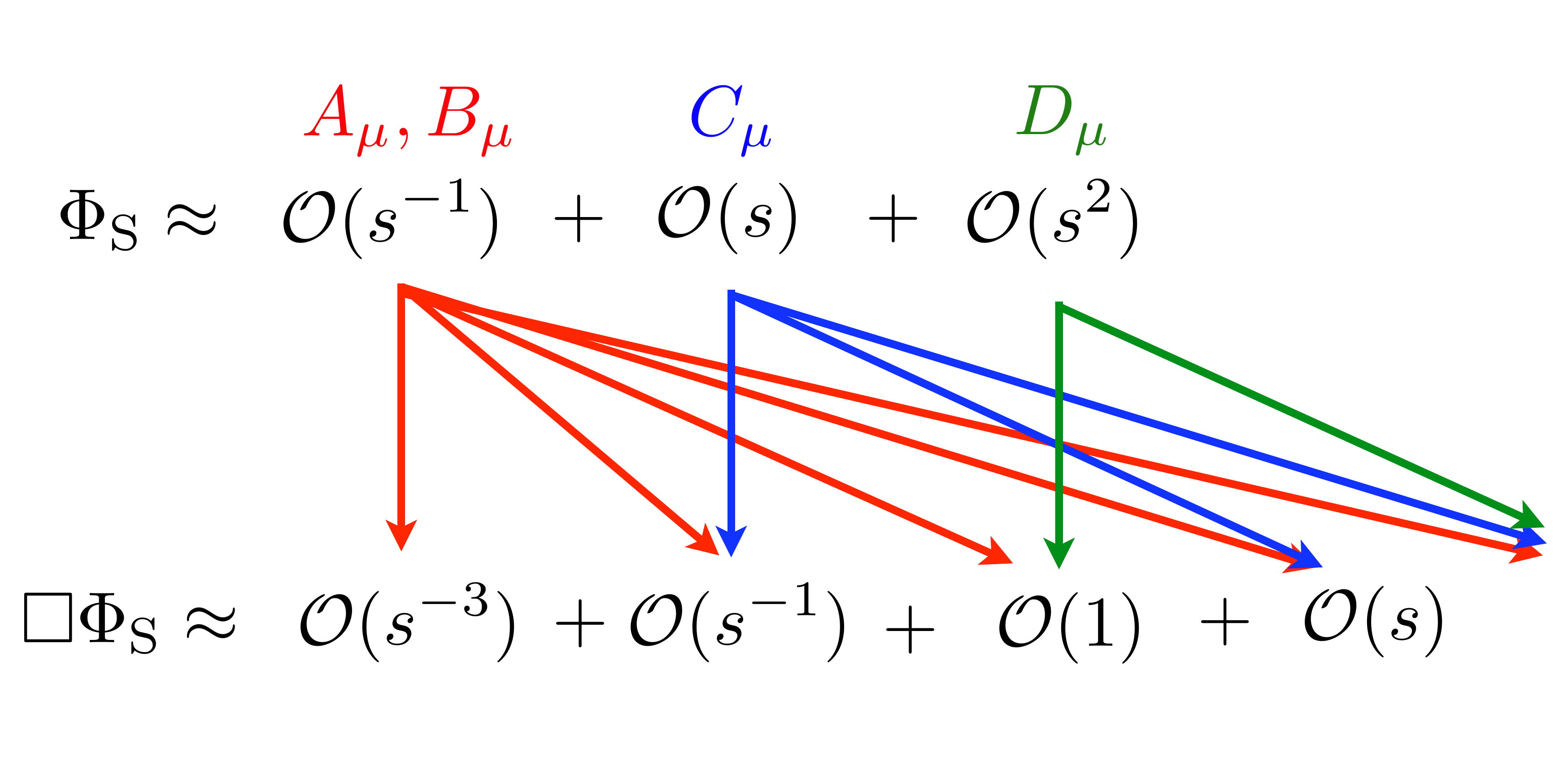}
\caption{Relation between order of approximation to the singular field and smoothness of the corresponding effective source. Also shown is the relation to the coefficients $A_\mu$, $B_\mu$, $C_\mu$, $D_\mu$, $\cdots$ in the large-$l$ expansion of the singular field used in the mode-sum regularization method \cite{Barack:Ori:2000}.}
\label{fig:ladder}
\end{figure}

We can clearly continue in this way (Fig.~\ref{fig:ladder}), producing a smoother source 
at each step. Taking this expansion to its logical conclusion, if we can calculate $\Phi_{\rm S}$ exactly, then we find that
\begin{equation}
S_{\rm eff}^{(\infty)} = 0
\end{equation}
and the self force comes purely from the boundary conditions. In practice, this would only require the computation of the expansion of $\Phi_{\rm S}$ to sufficiently high order to give an accurate numerical value in the region of interest. This may be difficult to impose in the window function approach, but in the world-tube approach the world-tube may be arbitrarily small and it may be possible. In that case, one would place a world-tube around the particle and then solve the system
\begin{eqnarray}
\mathcal{D} \Phi_1 = 0, \qquad \mathcal{D} \Phi_2 = 0,
\end{eqnarray}
where $\Phi_1$ is the field inside the tube and $\Phi_2$ is the full retarded field outside the tube. The self force then comes from applying the change of variables (i.e., boundary condition)
\begin{equation}
\Phi_2 |_{\rm W}= \Phi_1 |_{\rm W} + \Phi_{\rm S} |_{\rm W}
\end{equation}
across the world tube boundary $\rm W$. In this way, one may view the effective source as a correction for the fact that the singular field is not known exactly.

\section{Coordinate expressions and some practical considerations}
\label{sec:practical}
The previous section described the calculation of the singular field and effective source in a fully covariant manner. In practical applications, one needs to compute the singular field and effective source as a function of coordinate positions in a particular spacetime. A practical approach to doing so is to compute \emph{coordinate} expansions of the singular field and corresponding effective source. In this section, we develop such expansions and give example implementations in Schwarzschild and Kerr spacetimes. In doing so, we exploit insight from the covariant approach to simplify the calculations as much as possible.

\subsection{Coordinate expansion of singular field} \label{sec:PhiS-practical}
All terms in Eq.~\eqref{eq:PhiS-approx} may be written in terms of $\sigma_{\bar{a}}$ and local quantities at $\bar{x}$. In order to compute an explicit expression for a specific spacetime, it is convenient to expand $\sigma_{\bar{a}}$ in the coordinate separation between $x$ and $\bar{x}$ as follows \cite{Ottewill:Wardell:2008}:
\begin{enumerate}
\item Write $\sigma(x,\bar{x})$ as a formal coordinate series expansion about $\bar{x}$:
\begin{IEEEeqnarray}{rCl}
\label{eq:sigma-coord}
\sigma &=& \frac{1}{2} g_{\bar{\alpha} \bar{\beta}} \Delta x^{\bar{\alpha}} \Delta x^{\bar{\beta}} + A_{\bar{\alpha} \bar{\beta} \bar{\gamma}} \Delta x^{\bar{\alpha}} \Delta x^{\bar{\beta}} \Delta x^{\bar{\gamma}} + B_{\bar{\alpha} \bar{\beta} \bar{\gamma} \bar{\delta}} \Delta x^{\bar{\alpha}} \Delta x^{\bar{\beta}} \Delta x^{\bar{\gamma}} \Delta x^{\bar{\delta}} \nonumber \\
&&\qquad + C_{\bar{\alpha} \bar{\beta} \bar{\gamma} \bar{\delta} \bar{\epsilon}} \Delta x^{\bar{\alpha}} \Delta x^{\bar{\beta}} \Delta x^{\bar{\gamma}} \Delta x^{\bar{\delta}} \Delta x^{\bar{\epsilon}} + \cdots
\end{IEEEeqnarray}
where $\Delta x^{\bar{a}} \equiv x^a-\bar{x}^{\bar{a}}$ and where each of the coefficients is a function of $\bar{x}$ only and is symmetric in all indices.

\item Differentiate this expression at $\bar{x}$ to get 
\begin{align}
\sigma_{\bar{\alpha}} =& g_{\bar{\alpha} \bar{\beta}} \Delta x^{\bar{\beta}} + ( g_{\bar{\gamma} \bar{\beta}, \bar{\alpha}} + 3 A_{\bar{\alpha} \bar{\beta} \bar{\gamma}}) \Delta x^{\bar{\beta}} \Delta x^{\bar{\gamma}} + (A_{\bar{\beta} \bar{\gamma} \bar{\delta}, \bar{\alpha}} + 4 B_{\bar{\alpha} \bar{\beta} \bar{\gamma} \bar{\delta}}) \Delta x^{\bar{\beta}} \Delta x^{\bar{\gamma}} \Delta x^{\bar{\delta}} \nonumber \\
 & + (B_{\bar{\beta} \bar{\gamma} \bar{\delta} \epsilon, \bar{\alpha}} + 5 C_{\bar{\alpha} \bar{\beta} \bar{\gamma} \bar{\delta} \bar{\epsilon}}) \Delta x^{\bar{\beta}} \Delta x^{\bar{\gamma}} \Delta x^{\bar{\delta}} \Delta x^{\bar{\epsilon}} + \cdots
\end{align}

\item Use the identity $2 \sigma = \sigma_{\bar{\alpha}} \sigma^{\bar{\alpha}}$ to recursively determine the coefficients $A_{\bar{\alpha} \bar{\beta} \bar{\gamma}}$, $B_{\bar{\alpha} \bar{\beta} \bar{\gamma} \bar{\delta}}$.
\end{enumerate}
The result is a coordinate expansion of $\sigma_{\bar{a}}$ which may be substituted into \eqref{eq:PhiS-approx}. For the fourth order (i.e.\ $\mathcal{O}(\epsilon^2)$) approximation to the singular field, the coordinate expansion \eqref{eq:sigma-coord} must be computed to $\mathcal{O}\left[(\Delta x^{\bar{a}})^5\right]$ (i.e.\ the coefficients up to $C_{\bar{\alpha} \bar{\beta} \bar{\gamma} \bar{\delta} \bar{\epsilon}}$ must be determined). Since the coefficients are just functions of the metric and its partial derivatives at $\bar{x}$,  this calculation is easily achieved using a tensor software package such as GRTensorII \cite{GRTensor} or xCoba \cite{xTensor,xTensorOnline}. Rather than giving the full lengthy expressions, we present here only the leading two orders in the expansion in Schwarzschild spacetime to illustrate the structure:
\begin{IEEEeqnarray}{rCl}
\sigma(x,\bar{x}) &=&  \frac{\bar{r}}{2
   \left(\bar{r}-2M\right)}\Delta r^2+\frac{\bar{r}^2}{2}  \Delta \theta ^2
   +\frac{\bar{r}^2 }{2} \Delta \phi ^2 +
   \frac{2M-\bar{r}}{2 \bar{r}}\Delta t^2
   \nonumber \\
   &&
   -\frac{M}{2
   \left(\bar{r}-2M\right)^2}\Delta r^3+\frac{\bar{r}}{2}\Delta r\Delta \theta
   ^2 +\frac{\bar{r}}{2}\Delta r\Delta \phi ^2
   -\frac{M}{2
   \bar{r}^2}\Delta r \Delta t^2,
\end{IEEEeqnarray}
where the point $\bar{x}$ is assumed to lie in the equatorial plane (the spherical symmetry of Schwarzschild means that it is always possible to ensure this is the case). We provide a higher order expression for Kerr spacetime online \cite{EffSource-online}.

Next, we contract $\sigma_{\bar{a}}$ with the metric, Riemann tensor and four-velocity (all evaluated at $\bar{x}$) to get $\rbar$, $\sbar$ and Riemann terms such as $R_{u \sigma u \sigma}$. We then substitute these into Eq.~\eqref{eq:PhiS-approx} to obtain the coordinate expansion of $\tilde{\Phi}_{\rm S}$. In doing so, we only keep terms that contribute up to $\mathcal{O}(\epsilon^2)$. The first term in Eq.~\eqref{eq:PhiS-approx} is $\mathcal{O}(\epsilon^{-1})$ and so requires the coordinate expansion of $\sbar$ to order $\mathcal{O}\left[(\Delta x^{\bar{a}})^4\right]$ (equivalently, the expansion of $\sbar^2$ to order $\mathcal{O}\left[(\Delta x^{\bar{a}})^5\right]$). The second term is $\mathcal{O}(\epsilon)$ and requires $\rbar^2$, $\sbar^2$ and $R_{u \sigma u \sigma}$ to $\mathcal{O}\left[(\Delta x^{\bar{a}})^3\right]$. The third term is $\mathcal{O}(\epsilon^2)$ and requires $\rbar^2$, $\sbar^2$ and $R_{u \sigma u \sigma | u}$ to $\mathcal{O}\left[(\Delta x^{\bar{a}})^2\right]$ (the leading order) and $R_{u \sigma u \sigma | \sigma}$ to $\mathcal{O}(\mathcal{O}\left[(\Delta x^{\bar{a}})^3\right])$ (again, the leading order).  This results in an expression for the singular field which is valid to $\mathcal{O}\left[(\Delta x^{\bar{a}})^2\right]$ and has the general form
\begin{equation}
\label{eq:PhiS-coord1}
\tilde{\Phi}_{\rm S} = \frac{a_{(2)} + a_{(3)} + a_{(4)} + a_{(5)}}{(b_{(2)} + b_{(3)} + b_{(4)} + b_{(5)})^{3/2}}
\end{equation} 
where we use the notation $a_{(n)} = a_{\alpha_1 \cdots \alpha_n} \Delta x^{\alpha_1} \cdots \Delta x^{\alpha_n} $.

Although there is a clearly defined `true' singular field, in the effective source approach we may still view $\tilde{\Phi}_{\rm S}$ as merely a computational tool with a certain degree of flexibility in choosing its particular form. Indeed, this coordinate approximation to the singular field is not unique -- the only requirement it must satisfy is that it matches the `true' singular field to a prescribed order -- and it may therefore be replaced with any other expression which agrees with it to the desired order.

The expression in the denominator of Eq. \eqref{eq:PhiS-coord1} is undesirable because it leads to long calculations, particularly when computing the derivatives required for the effective source corresponding to this choice of singular field. More importantly, the roots of this denominator are singularities in $\tilde{\Phi}_{\rm S}$ and, potentially, in $S_{\rm eff}$. Since it is a power of a fifth-order polynomial, the denominator will have roots different from the trivial one, $\Delta x^\alpha =0$, which represents the worldline of the particle. As a result, the effective source will have undesirable divergences at certain coordinate locations. (Note that it is $C^0$ at the location of the particle). Moreover, on any given time slice, the precise location of these singularities will depend sensitively on the position and four-velocity of the particle (on which the coefficients $b_{\alpha_1\alpha_2\ldots\alpha_n}$ depend). The presence of these extra singularities is purely an artifact of using a truncated series expansion to approximate $\sbar$; the exact $\sbar$ increases monotonically away from the particle. This becomes problematic for any numerical application. 

For these reasons, it is advantageous to modify the singular field produced from the above described procedure. Noting that $\Delta x^a = \mathcal{O}(\epsilon)$, we re-expand the coordinate expansion of $\Phi_{\rm S}$ about $\epsilon=0$. In practice this is most easily achieved by introducing an explicit factor of $\epsilon$ into the coordinate distances, $\Delta x^a \to \epsilon \Delta x^a$, expanding about $\epsilon=0$ (to $\mathcal{O}(\epsilon^2)$ for the fourth order singular field) and reading off the coefficient of each power of $\epsilon$. The result is an approximation to the singular field of the form
\begin{equation}
\label{eq:PhiS-coord2}
\tilde{\Phi}_{\rm S} = \frac{c_{(6)} + c_{(7)} + c_{(8)} + c_{(9)}}{(b_{(2)})^{7/2}},
\end{equation}
with a new denominator whose roots are much more manageable.
In particular, the re-expansion leaves only the $O(\Delta x^2)$ terms, or those that are quadratic in the coordinate displacements, of the 
original denominator. 
From Eq. \eqref{eq:PhiS-approx}, we see that only $\sbar$ 
appears in the denominator, whose quadratic dependence on the coordinate displacement is simply 
\begin{equation}
\sbar^2 = g_{\bar{\alpha}\bar{\beta}}\Delta x^{\bar{\alpha}}\Delta x^{\bar{\beta}}+ (u_{\bar{\alpha}}\Delta x^{\bar{\alpha}})^2 + O(\Delta x^3)
\end{equation}
The second term is manifestly positive except at the location of the particle where it vanishes. The first term is not 
necessarily positive and may still potentially result in a vanishing denominator, in general. However, if in some 
coordinate system one chooses to associate the field point, $\bar{x}$, with the particle position, $x$, so that 
they always share a common time coordinate (that is, $t=\bar{t}$), then we have 
$\sbar^2(t=\bar{t})= g_{\bar{i}\bar{j}}\Delta x^{\bar{i}}\Delta x^{\bar{j}}+ (u_{\bar{\alpha}}\Delta x^{\bar{\alpha}})^2$.
Now, since $g_{\bar{i}\bar{j}}$ is a purely spatial metric, its eigenvalues are all positive, so that
the first term is unconditionally positive-definite and only vanishes at the location of the particle. (See Appendix \ref{sec:s2positive} 
for an explicit demonstration in the case of Schwarzschild coordinates). 
Thus, with a re-expansion of the denominator we achieve a simplification and, more importantly, we are also able to avoid the non-worldline 
singularities in the Haas-Poisson expression for the singular field given in Eq.~\eqref{eq:PhiS-coord1}. The latter feature is essential for, say, 
a robust (3+1) application of the effective source approach. It is important to remember that to guarantee this, $x$ and $\bar{x}$ 
must be on the same $t$-hypersurface, where $t$ is the time coordinate in the specific coordinate system 
chosen to express $\tilde{\Phi}_{\rm S}$. 

\subsection{Periodicity of the singular field} \label{sec:periodicity}
Although not strictly necessary, in spacetimes with axial symmetry it may be desirable to have
an approximation to the singular field which is periodic in the azimuthal coordinate. There is
no guarantee that that will be the case for the expansions \eqref{eq:PhiS-coord1} and
\eqref{eq:PhiS-coord2}; in fact there is not even any guarantee that
$\tilde{\Phi}_{\rm S}(\Delta \phi = \pi) = \tilde{\Phi}_{\rm S}(\Delta \phi = -\pi)$, i.e.\ that
it is continuous across $\Delta \phi = \pm \pi$.

Barack and Golbourn \cite{Barack:Golbourn:2007} explicitly enforce periodicity by making
the substitution $\Delta \phi^2 = 2(1-\cos \Delta \phi) + \mathcal{O}(\Delta \phi^4)$. This was
extended to higher order in Ref.~\cite{dolan-etal:11} by making use of expressions
involving $\cos(\Delta\phi)$ and $\cos(2 \Delta\phi)$. However, both of these previous works
only required replacements for even powers of $\Delta \phi$. In general odd powers
of $\Delta \phi$ can (and do) also appear.

Among the infinitely many ways in which periodicity may be enforced
for both odd and even powers, not all approaches are equal. For example, using replacements
involving $\sin n \Delta \phi$ proves to be a poor choice; since $\sin n\Delta\phi = 0$
at $\Delta\phi = \pm \pi$ for any integer~$n$ such replacements may lead to the denominator of
\eqref{eq:PhiS-coord1} or \eqref{eq:PhiS-coord2} vanishing if $\Delta\phi = \pm \pi$ lies within
the worldtube~(Sec.~\ref{sec:worldtube}) or the window function's
region of support~(Sec.~\ref{sect:window-fn}). Unfortunately, it is easy to see that no
alternative choice for the
functions used to replace odd powers of $\Delta\phi$ can avoid such
extra zeros.  That is, denoting the replacement for $\Delta\phi^n$
by $f_n(\Delta\Phi)$, for any odd~$n$ the function $f_n$ \emph{must}
have at least one zero somewhere in $\Delta\phi \in (0,2\pi)$.\footnote{
	 To see this, suppose that $n$ is a (positive) odd integer.
	 $f_n$ must clearly satisfy the following properties (among others):
	 \begin{enumerate}
	 \item\label{item-f-n-continuous}
		$f_n$ is continuous
	 \item\label{item-f-n-approx-x-for-small-|x|}
		$f_n(x) \approx x^n$ for small $|x|$  (indeed, $f(x) = x^n + \mathcal{O}(x^m)$,
		where $m$ is the highest power of $x$ appearing in the expansion of the singular field)
	 \item\label{item-f-n-periodic}
		$f_n(x + 2k\pi) = f_n(x)$ for any integer~$k$
	 \end{enumerate}
	 Property~\ref{item-f-n-approx-x-for-small-|x|} implies that
	 if $n$~is odd, $f_n > 0$ for small positive $\Delta\phi$, and
	 $f_n < 0$ for small negative $\Delta\phi$.
	 Property~\ref{item-f-n-periodic} then implies that
	 $f_n < 0$ for $\Delta\phi$ slightly less than $2\pi$.
	 Property~\ref{item-f-n-continuous} and the intermediate
	 value theorem then imply that $f_n$ must have a zero
	 somewhere between $\Delta\phi = 0$ and $\Delta\phi = 2\pi$.
	 }

Given the two criteria: (i)
$\tilde{\Phi}_{\rm S}(\Delta \phi = \pi) = \tilde{\Phi}_{\rm S}(\Delta \phi = -\pi)$
and (ii) $\tilde{\Phi}_{\rm S}(\Delta \phi = \pi)$ is finite, we therefore propose a particular choice
which satisfies both requirements and which has other practical advantages. We introduce the angular variables
\begin{equation}
Q = 	\sin \thalf\Delta\phi, \qquad R =	\sin \Delta\phi
\end{equation}
and rewrite \emph{even} powers of $\Delta \phi$ in terms of $Q$ and \emph{odd} powers in terms
of $Q$ and $R$. This is easily achieved by expanding $\Delta \phi = 2 \arcsin Q$ and
$\Delta \phi = \arcsin R$ for small $Q$ and $R$, and making use of the identity $R^2=4Q^2(1-Q^2)$ to give
\begin{gather}
\Delta \phi \approx R\Big(1 + \frac{2}{3} Q^4 + \frac{8}{15} Q^5 \Big), \quad
\Delta \phi^2 \approx 4Q^2 + \frac{4}{3} Q^4, \quad
\Delta \phi^3 \approx 4 R Q^2\Big( 1+Q^2\Big), \qquad
\Delta \phi^4 \approx 16 Q^4, \qquad
\Delta \phi^5 \approx R^5,
\label{eq:phi5subs}
\end{gather}
where terms of $\mathcal{O}(\Delta \phi^6)$ and higher have been neglected.
Not only does this replacement satisfy both criteria mentioned above, it also leads
to relatively compact formulas for the
partial derivatives 
\begin{equation}
\partial_{\phi} = \frac{R}{4Q} \partial_Q + (1-2Q^2) \partial_R, \qquad
\partial_{\phi\phi} = \frac14 (1-Q^2) \partial_{QQ} - \frac14 Q \partial_Q + \frac{R}{2Q}(1-2Q^2) \partial_{QR} - R \partial_R + (1-2Q^2)^2 \partial_{RR}
\end{equation}
which appear in the
wave operator (used when calculating the effective source).
Moreover, this choice of variables has the
subtle advantage of lending itself to minimal sensitivity to round-off effects close to
the particle (see Sec.~\ref{sec:cancellation} for an explanation of why this is important).
For small $\Delta \phi$, the substitutions of Refs.~\cite{Barack:Golbourn:2007}
and \cite{dolan-etal:11} are sensitive to numerical round-off, whereas this is not the case for our
$(Q, R)$ scheme.

\subsection{Calculation of the effective source}
With an approximation to the singular field at hand, we must now calculate a corresponding
effective source. Before proceeding with the calculation, we will briefly mention some
issues which one must be cognizant of.

In general, the calculation of an effective source requires the computation of derivatives of $\Phi_{\rm S}(x)$.
When calculating these derivatives, one generally needs to be careful to take account of the fact
that $x'$ and $x''$ vary with $x$ since they must remain linked by a null geodesic
\cite{Poisson:2003}. Fortunately, in the approximation $\tilde{\Phi}_{\rm S}$ of
Eq.~\eqref{eq:PhiS-approx}, by writing everything in terms of $\sigma_{\bar{a}}$, this
dependence is made explicit in terms of the arbitrary point $\bar{x}$ which does \emph{not}
depend on $x$. However, since this dependence is only given as an expansion in $\epsilon$, it
is an approximation which is only strictly valid in the limit $\epsilon \to 0$. For example, it
is possible (and likely) that \eqref{eq:PhiS-approx} differs from the `true' singular field at
$\mathcal{O}(\epsilon^3)$, yet in the limit $\epsilon \to 0$ they agree. Similarly, the
corresponding effective source has the correct value (i.e.\ $0$) at $\epsilon=0$, but contains
differences from the `true' effective source at $\mathcal{O}(\epsilon)$. One must be
careful to account for this when computing an effective source.

In computing a covariant approximation to the effective source in Sec.~\ref{sec:effsrc-approx}, we made use of identities such as $\sigma^{\bar{\alpha}}\sigma_{\bar{\alpha}} = 2 \sigma$ and $\sigma^{\alpha}\sigma_{\alpha} = 2 \sigma$. Furthermore, we re-expressed higher derivatives of $\sigma$ in terms of their covariant expansion about $\bar{x}$. However, once coordinate expansions are introduced, these identities and expansions are no longer exact -- they are only valid up to the order of the coordinate expansion. In computing the singular field, this is not an issue since we are only interested in computing the value of the self-force at the particle, in which case the errors vanish. Unfortunately, the effective source is required not just at the particle, but also in a region surrounding the particle where the errors are no longer zero. It is therefore not possible to make use of these simplifications when calculating a coordinate effective source (at least not without taking care that $\Phi_{\rm S}$ and its derivative evaluated at the particle are unchanged).

With these issues in mind, there are now two choices on how to proceed with computing $S_{\rm eff} = -\Box \tilde{\Phi}_{\rm S}$. We will investigate both of these in turn in the following sections.

\subsubsection{World-tube method}
\label{sec:worldtube}
Barack and Golbourn \cite{Barack:Golbourn:2007} propose a precise method for computing an effective source. They introduce a world-tube around the particle. Inside the world-tube one solves for $\tilde{\Phi}_{\rm R}$ and outside one solves for $\Phi_{\rm ret}$, which is now a solution of the \emph{homogeneous} wave equation in this region. By imposing the boundary condition $\Phi_{\rm ret} = \tilde{\Phi}_{\rm R}+ \tilde{\Phi}_{\rm S}$ one can ensure that the system as a whole is consistent.

They look for a `puncture' field - an approximation to the singular field which depends only on the spatial position of the field point, with all time dependence encapsulated in the particle motion,
\begin{equation}
\Phi_{\rm P} (x^i, x^{\bar{j}}(t), u^{\bar{a}}(t)).
\end{equation}
In doing so, they effectively fix $\bar{t} = t$, i.e.\ fixing $\bar{x}$ to depend on $x$ in the sense that their time coordinates are equal. Recall that $\bar{x}$ is arbitrary and does not have any required dependence on $x$. Their choice is therefore valid and consistent with the singular field computed in Sec.~\ref{sec:PhiS-practical}.
In particular, their choice of puncture function
\begin{equation}
\Phi_{\rm P} = \frac{1}{\sqrt{(g_{i j} + u_i u_j) \Delta x^i \Delta x^j}},
\end{equation}
corresponds exactly to the first order singular field given here (with $\bar{t}=t$). 
In Ref.~\cite{Barack:Golbourn:Sago:2007}, Barack, Golbourn and Sago proposed an improved
puncture function, which again is equivalent to the second order singular field given here.
Computing higher order puncture functions is straightforward: one takes the expansions
\eqref{eq:PhiS-coord1} or \eqref{eq:PhiS-coord2} at the desired order and sets $\Delta t = 0$. For
example, a fourth order puncture function for a particle in circular equatorial geodesic motion
around a Kerr black hole is given explicitly in Sec.~\ref{sec:Kerr-circular}. Dolan and Barack
\cite{Dolan:Barack:2010} have recently made use of a similar fourth-order puncture computed in this
way to calculate the self-force on a particle in a circular geodesic orbit about a Schwarzschild
black hole and Dolan, Barack and Wardell \cite{dolan-etal:11} extended this to the Kerr case.

Given this puncture field, the computation of an associated effective source is straightforward.
One simply calculates an expression for the wave operator in the coordinates in which
$\Phi_{\rm P}$ is given and applies this wave operator to $\Phi_{\rm P}$, noting that spatial
derivatives act only on $x^i$, while time derivatives act only on $x^{\bar{j}}(t)$
and $u^{\bar{a}}(t)$.

\subsubsection{Window function method}
\label{sect:window-fn}
In a numerical $3+1$ evolution code, it is most straightforward to solve for $\Phi_{\rm R}$ everywhere, requiring $S_{\rm eff}$ everywhere on a 3D spatial slice. This would be problematic wherever $\Phi_{\rm S}$ is either not defined or where its series expansion diverges. Vega and Detweiler \cite{Vega:Detweiler:2008} propose a solution which involves the introduction of a window function, $W(r)$, which smoothly transitions from a value of $1$ at the source to $0$ far away. In effect, one is then solving for $\Phi_{\rm R}$ near the particle and for $\Phi_{\rm ret}$ far from the particle. For this to provide a consistent $\mathcal{C}^0$ source, we must impose a restriction on $W$: at the particle it must be $1$ and at least its first three derivatives must be $0$. Having introduced this window function, the effective source is then given by 
\begin{equation}
S_{\rm eff} = \mathcal{D} (W \tilde{\Phi}_S).
\end{equation}
In \cite{Vega:2009}, Vega et al. use an expression for $S_{\rm eff}$ which is $\mathcal{C}^0$ (i.e.\ continuous but not differentiable), limiting the convergence of their finite differencing scheme despite their use of $8$-th order spatial differencing. A smoother source would be advantageous in that it would give a higher convergence order without the need to construct a more complicated finite differencing scheme to deal with the non-smoothness of the source. In \cite{diener-etal:12b} this work was extended to include the back reaction from the self-force into the evolution. This 
latter calculation made use of the re-expanded singular field described above, evaluated in Kerr-Schild coordinates with the choice $\bar{t}_{\rm KS}=t_{\rm KS}$.

\subsubsection{Evaluation of the effective source very close to the particle}
\label{sec:cancellation}
Severe round-off errors may be incurred when evaluating the effective source very close to the particle. Applying the wave operator 
to the singular field results in many terms that scale as $O(\epsilon^{-3})$, which evaluate to large quantities as 
$\epsilon \rightarrow 0$. However, we show in the analysis of Sec. \ref{sec:effsrc-approx} that, at the order of our present 
approximation to the singular field, all of these terms cancel to leave an over-all effective source that 
scales as $O(\epsilon)$. As was already pointed out in \cite{vega-etal:11}, this is a prototypical example of 
catastrophic cancellation that is often encountered in numerical work. There are two solutions
to this problem which have been found to work well. We describe each approach in detail
below and note that the choice of which scheme to use is
dependent on the problem at hand. For simpler configurations (with more manageable expressions) a
series approximation may be appropriate, whereas in cases where more
unwieldy expressions appear it may be more straightforward to use numerical interpolation.

\emph{Series approximation close to the particle}

\begin{figure}
\includegraphics[width=6cm]{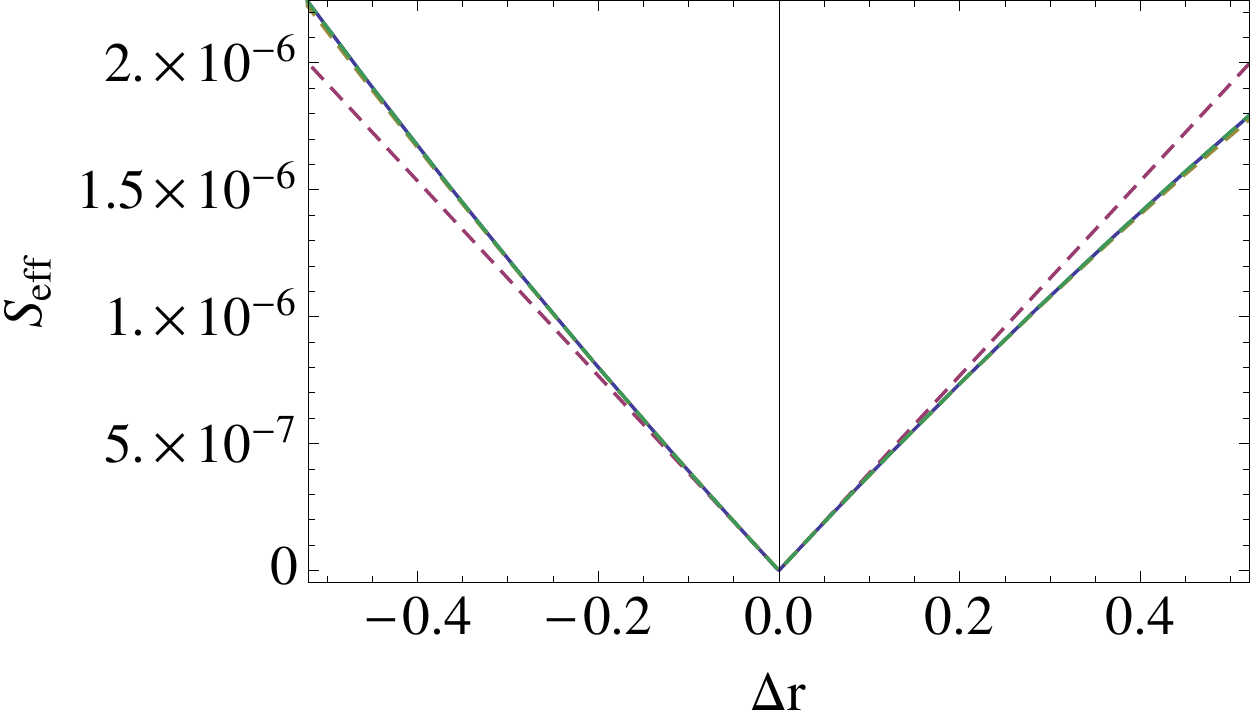}
\caption{Coordinate series expansion approximation to the effective source close to the
particle. The exact effective source (solid blue line) is well approximated in the region
$\Delta r \lesssim 0.05M$ by the first term in its series expansion (dashed purple line).
Including the second (dashed gold line) and third (dashed green line) terms further improves the
agreement for points farther from the particle, with the curves being almost
indistinguishable from the exact curve.}
\label{fig:source-approx}
\end{figure}

Given that cancellation is only an issue for points very close to the particle (typically
at a distance of $\lesssim 0.05M$), it is reasonable to replace the full effective source
in this region with an approximation which is valid for points a small distance
from the particle. In particular, by replacing the full effective source with its series
expansion\footnote{Although our approximation to the singular field is already written as a truncated series
expansion, the effective source which is computed by applying the wave operator to it is not.
For example, as can be seen from Eqs.~\eqref{eq:box-schw} and \eqref{eq:box-kerr} there are several
terms which depend on the location where the source is being evaluated and which are not written
explicitly as series expansions.}, one obtains an expression for the effective source of the form
\begin{equation}
\label{eq:src-approx}
\tilde{S}_{\rm eff} = \frac{d_{(12)} + d_{(13)} + d_{(14)}}{(b_{(2)})^{11/2}} + \mathcal{O}(\epsilon^4).
\end{equation}
This expression is manifestly $\mathcal{O}(\epsilon)$, with all divergent terms having been
cancelled analytically. For the small region where catastrophic cancellation arises, it is
sufficient to take only the first term, $\frac{d_{(12)}}{(b_{2})^{11/2}}$, which is
$\mathcal{O}(\epsilon)$. The inclusion of subsequent terms would only be necessary
if an approximation was needed in a much larger region (see Fig.~\ref{fig:source-approx}).
There is a potential disadvantage to this scheme, however, in that it involves the
evaluation of the twelfth order
polynomial, $d_{(12)}$, which may be quite computationally expensive. Fortunately, since this is 
to be implemented only in a very small region around the particle, the overall computational 
burden this adds is likely to be minimal. 

\emph{Interpolation close to the particle}

Another solution to the catastrophic cancellation problem is 
to compute the effective source via interpolation.  
We take advantage of two important facts: (1) $S_{\rm eff}=0$ on the worldline $\gamma$, and (2) $S_{\rm eff}$ is 
smooth everywhere except on $\gamma$, where it is just $C^0$ . The first fact gives us an exact data point 
for the effective source, while the second justifies an assumption that interpolation might be sufficient.
We identify a small region $\mathcal{R}$ 
around the particle location outside of which the effective source is computed reasonably well. If the effective source
is required inside $\mathcal{R}$, say at $y^i$, then it is first evaluated at 
selected points along a ``coordinate ray" outside this region. Using these values and $S_{\rm eff}(x^i = \bar{x}^i)  = 0$, 
where $\bar{x}^i$ are the spatial cooordinates of the location of the particle, we then interpolate to $y^i$.
(All coordinates here are purely spatial in compliance with the restriction mentioned in Sec. \ref{sec:PhiS-practical}:
when evaluating the effective source, all field points must be on the same $t$-hypersurface as the particle.)

More concretely, consider $S(\lambda) := S_{\rm eff}(x^i(\lambda))$ as a function of $\lambda$, along the coordinate ray
given by $x^i(\lambda) = \bar{x}^i + \lambda(y^i-\bar{x}^i)$. If $y^i \in \mathcal{R}$, then to compute 
$S_{\rm eff}(y^i) = S(\lambda =1)$, we interpolate using a few evaluations of $S(\lambda_j)$ [such that
$x^i(\lambda_j) \notin \mathcal{R}$] and $S(\lambda=0) = 0$. 

Obviously, there is considerable freedom in how to implement a specific interpolation scheme and in what to choose for the 
size of the interpolation region $\mathcal{R}$. The results reported in \cite{diener-etal:12a} appear to be 
very robust with respect to the various choices we have tried.

\subsection{Specific schemes}
There are three commonly applied approaches which may be used for solving the wave equation for the regularized field. These methods solve for the regularized field in $1+1$, $2+1$ and $3+1$ dimensions, eliminating the other dimensions through a decomposition in suitable basis functions. In the $1+1$D approach, the spherical harmonic basis is chosen and the decomposition is done into $l$ and $m$ modes, while in $2+1$ dimensions the decomposition is only done into $m$ modes. There is a trade off between having to evolve a field in higher dimension (and all the difficulties of poor scaling with resolution that goes with it) and requiring the calculation of a large number of modes. It remains to be seen which is the better choice; the conclusion will most likely depend on the particular configuration under consideration. Nevertheless, the calculation of the singular field and effective source proceeds in the same way. Both are calculated as $3+1$ dimensional quantities using the methods described in the previous sections. In the $2+1$D $m$-mode scheme, they are then decomposed into $m$ modes by performing an integration over the azimuthal coordinate, $\phi$:
\begin{equation}
\tilde{\Phi}_{\rm S~(m)} = \int_{-\pi}^{\pi} \tilde{\Phi}_{\rm S}(\Delta r, \Delta \theta, \Delta \phi, \Delta t) e^{-i m \phi} d \phi \qquad S_{\rm eff~(m)} = \int_{-\pi}^{\pi} S_{\rm eff}(\Delta r, \Delta \theta, \Delta \phi, \Delta t) e^{-i m \phi} d \phi.
\end{equation}
For the $1+1$D $l,m$-mode scheme, a second integration is performed over the inclination angle $\theta$:
\begin{IEEEeqnarray}{rCl}
\tilde{\Phi}_{\rm S~(l,m)} &=& \int_{-\pi}^{\pi} \int_0^\pi \tilde{\Phi}_{\rm S}(\Delta r, \Delta \theta, \Delta \phi, \Delta t) Y^*_{l,m} (\theta, \phi) d\theta d \phi \nonumber \\
 S_{\rm eff~(l,m)} &=& \int_{-\pi}^{\pi} \int_0^\pi S_{\rm eff}(\Delta r, \Delta \theta, \Delta \phi, \Delta t) Y^*_{l,m} (\theta, \phi)  d\theta d \phi.
\end{IEEEeqnarray}
In practice it may be most straightforward to do the integration numerically. As a result, the calculation of the singular field and effective source may dominate the runtime of a $1+1$D or $2+1$D code.

\section{Examples}
\label{sec:examples}
In this section, we give examples of the singular field and effective source in some specific scenarios. We consider in detail the case of a scalar charge undergoing circular, equatorial, geodesic motion in both Schwarzschild and Kerr spacetimes. Note, however, that the methods developed here do not depend on the symmetries present in these configurations. They are equally effective in other spacetimes and for generic geodesic motion. For these more generic configurations, the results are most easily given in electronic form. For this reason, we provide expressions for more generic configurations online \cite{EffSource-online} and give here only explicit examples for simple configurations along with plots for more generic configurations.

\subsection{Circular geodesic in Schwarzschild}
\label{sec:Schw-circular}
Given the Schwarzschild metric in standard coordinates,
\begin{equation}
ds^2 = -\left(1 - \frac{2M}{r}\right) dt^2 + \left(1 - \frac{2M}{r}\right)^{-1} dr^2 + r^2 d\theta^2+ r^2\sin^2\theta d\phi^2,
\end{equation}
we follow the prescription of Sec.~\ref{sec:practical} to obtain a fourth order approximation to the singular field of the kind given in \eqref{eq:PhiS-coord2}. In general, this will be a function of the field point, $x = (r, \theta, \phi, t)$, the world-line point $\bar{x} = (\bar{r}, \bar{\theta}, \bar{\phi}, \bar{t})$ and the particle four-velocity $u^a = (u^r, u^\theta, u^\phi, u^t)$. We may use the spherical symmetry of the spacetime to enforce that the motion lies in the equatorial plane, i.e.\ $\bar{\theta} = \pi/2$, $u^\theta = 0$. In order to obtain sufficiently compact expressions to be given here, we make the further assumption that the motion is circular, i.e.\ \cite{Chandrasekhar}
\begin{equation}
\bar{r} = \text{constant} \qquad u^r = 0 \qquad u^\phi = \frac{1}{\bar{r}} \sqrt{\frac{M}{\bar{r}-3M}} \quad u^t = \sqrt{\frac{\bar{r}}{\bar{r}-3M}}.
\end{equation}
Finally, we use the freedom in the choice of $\bar{t}$ to set the field point and world-line point to be at the same coordinate time, i.e.
\begin{equation}
\bar{t} = t \qquad \bar{\phi} = \Omega t,
\end{equation}
where
\begin{equation}
\Omega = \sqrt{\frac{M}{\bar{r}^3}}
\end{equation}
is the orbital frequency.
Combining everything, we obtain a fourth order approximation to the singular field of a scalar charge on a circular, equatorial orbit around a Schwarszchild black hole:
\begin{equation}
\label{eq:PhiS-Schw}
\tilde{\Phi}_{\rm S}^{(4)}(r, \theta, Q, t) = \frac{\sum_{i,j,k = 0}^{i+j+k\le9} a_{ijk} \Delta r^i \Delta \theta^j Q^k}{\left(\sum_{i,j,k = 0}^{i+j+k\le2} b_{ijk} \Delta r^i \Delta \theta^j Q^k\right)^{7/2}}
\end{equation}
where $\Delta r = r-\bar{r}$, $\Delta \theta = \theta - \pi/2$, $Q = \sin\big(\frac12(\phi - \Omega t)\big)$ and 
where the non-zero coefficients, $a_{ijk}$ and $b_{ijk}$ are functions of the orbital radius, $\bar{r}$, and are given by
\begin{gather}
a_{006} = -\frac{64 \bar{r}^6 (2 M-\bar{r})^3}{(\bar{r}-3 M)^3}, \quad
a_{024}=\frac{48 \bar{r}^6 (\bar{r}-2 M)^2}{(\bar{r}-3 M)^2}, \quad
a_{042}=\frac{12 \bar{r}^6 (\bar{r}-2 M)}{\bar{r}-3 M}, \quad
a_{060}=\bar{r}^6, \quad
a_{204}=\frac{48 \bar{r}^5 (\bar{r}-2 M)}{(\bar{r}-3 M)^2}, \nonumber \\
a_{222}=-\frac{24 \bar{r}^5}{3 M-\bar{r}}, \quad
a_{240}=-\frac{3 \bar{r}^5}{2 M-\bar{r}}, \quad
a_{402}=\frac{12 \bar{r}^4}{6 M^2-5 M \bar{r}+\bar{r}^2}, \quad
a_{420}=\frac{3 \bar{r}^4}{(\bar{r}-2 M)^2}, \quad
a_{600}=\frac{\bar{r}^3}{(\bar{r}-2 M)^3}, \nonumber \\
a_{106}=-\frac{32 \bar{r}^5 (M-\bar{r}) (\bar{r}-2 M)^2}{(3 M-\bar{r})^3}, \quad
a_{124}=-\frac{8 \bar{r}^5 \left(8 M^2-10 M \bar{r}+3 \bar{r}^2\right)}{(\bar{r}-3 M)^2}, \quad
a_{142}=\frac{2 \bar{r}^5 (3 \bar{r}-5 M)}{3 M-\bar{r}}, \quad
a_{160}=-\frac{\bar{r}^5}{2}, \nonumber \\
a_{304}=\frac{8 \bar{r}^4 (3 M-2 \bar{r})}{(\bar{r}-3 M)^2}, \quad
a_{322}=\frac{8 \bar{r}^4}{3 M-\bar{r}}, \quad
a_{340}=\frac{\bar{r}^4 (5 M-2 \bar{r})}{2 (\bar{r}-2 M)^2}, \quad
a_{502}=-\frac{2 \bar{r}^3}{(\bar{r}-2 M)^2}, \quad
a_{520}=-\frac{\bar{r}^3 (\bar{r}-4 M)}{2 (\bar{r}-2 M)^3},\nonumber \\
a_{700}=\frac{M \bar{r}^2}{2 (\bar{r}-2 M)^4}, \quad
a_{008}=-\frac{32 M \bar{r}^5 (2 M-\bar{r})^3}{(3 M-\bar{r})^3}, \quad
a_{026}=-\frac{16 \bar{r}^5 (\bar{r}-2 M)^2 \left(5 M^2-4 M \bar{r}+\bar{r}^2\right)}{(3 M-\bar{r})^3}, \nonumber \\
a_{044}=\frac{2 \bar{r}^5 \left(222 M^4-459 M^3 \bar{r}+346 M^2 \bar{r}^2-112 M \bar{r}^3+13 \bar{r}^4\right)}{3 (\bar{r}-3 M)^3}, \quad
a_{062}=\frac{\bar{r}^5 \left(-30 M^3+57 M^2 \bar{r}-29 M \bar{r}^2+4 \bar{r}^3\right)}{3 (\bar{r}-3 M)^2}, \nonumber \\
a_{080}=-\frac{\bar{r}^5 \left(6 M^2-9 M \bar{r}+\bar{r}^2\right)}{72 M-24 \bar{r}}, \quad
a_{206}=\frac{8 \bar{r}^4 \left(-30 M^3+35 M^2 \bar{r}-16 M \bar{r}^2+3 \bar{r}^3\right)}{(\bar{r}-3 M)^3}, \nonumber \\
a_{224}=\frac{2 \bar{r}^4 \left(37 M^2-40 M \bar{r}+13 \bar{r}^2\right)}{(\bar{r}-3 M)^2}, \quad
a_{242}=\frac{\bar{r}^4 \left(294 M^3-498 M^2 \bar{r}+259 M \bar{r}^2-41 \bar{r}^3\right)}{6 (2 M-\bar{r}) (\bar{r}-3 M)^2}, \nonumber \\
a_{260}=\frac{\bar{r}^4 \left(48 M^2-57 M \bar{r}+11 \bar{r}^2\right)}{24 \left(6 M^2-5 M \bar{r}+\bar{r}^2\right)}, \quad
a_{404}=\frac{2 \bar{r}^3 \left(-65 M^3+74 M^2 \bar{r}-26 M \bar{r}^2+3 \bar{r}^3\right)}{(2 M-\bar{r}) (3 M-\bar{r})^3}, \nonumber \\
a_{422}=\frac{-50 M^3 \bar{r}^3+72 M^2 \bar{r}^4-30 M \bar{r}^5+4 \bar{r}^6}{\left(6 M^2-5 M \bar{r}+\bar{r}^2\right)^2}, \quad
a_{440}=\frac{5 \bar{r}^3 \left(9 M^2-10 M \bar{r}+2 \bar{r}^2\right)}{24 (\bar{r}-2 M)^3}, \nonumber \\
a_{602}=\frac{M \bar{r}^2 \left(31 M^2-37 M \bar{r}+8 \bar{r}^2\right)}{2 (2 M-\bar{r})^3 (\bar{r}-3 M)^2}, \quad
a_{620}=\frac{M \bar{r}^2 \left(19 M^2-21 M \bar{r}+4 \bar{r}^2\right)}{8 (3 M-\bar{r}) (\bar{r}-2 M)^4}, \nonumber \\
a_{800}=\frac{M^2 \bar{r} (\bar{r}-M)}{4 (2 M-\bar{r})^5 (3 M-\bar{r})}, \quad
a_{108}=\frac{16 M \bar{r}^4 (\bar{r}-2 M)^2 \left(29 M^3-25 M^2 \bar{r}+3 M \bar{r}^2+\bar{r}^3\right)}{(3 M-\bar{r})^5}, \nonumber \\
a_{126}=\frac{4 \bar{r}^4 (2 M-\bar{r}) \left(97 M^4-86 M^3 \bar{r}+27 M^2 \bar{r}^2-8 M \bar{r}^3+2 \bar{r}^4\right)}{(\bar{r}-3 M)^4}, \nonumber \\
a_{144}=-\frac{\bar{r}^4 \left(312 M^4-351 M^3 \bar{r}+193 M^2 \bar{r}^2-73 M \bar{r}^3+13 \bar{r}^4\right)}{3 (\bar{r}-3 M)^3}, \quad
a_{162}=\frac{\bar{r}^4 \left(54 M^3-51 M^2 \bar{r}+31 M \bar{r}^2-8 \bar{r}^3\right)}{12 (\bar{r}-3 M)^2}, \nonumber \\
a_{180}=\frac{\bar{r}^5 (3 M+\bar{r})}{48 (3 M-\bar{r})}, \quad
a_{306}=-\frac{4 \bar{r}^3 \left(139 M^4-163 M^3 \bar{r}+78 M^2 \bar{r}^2-27 M \bar{r}^3+5 \bar{r}^4\right)}{(\bar{r}-3 M)^4}, \nonumber \\
a_{324}=\frac{\bar{r}^3 \left(-357 M^4+434 M^3 \bar{r}-230 M^2 \bar{r}^2+78 M \bar{r}^3-13 \bar{r}^4\right)}{(2 M-\bar{r}) (3 M-\bar{r})^3}, \nonumber \\
a_{342}=-\frac{\bar{r}^3 \left(732 M^4-1074 M^3 \bar{r}+677 M^2 \bar{r}^2-239 M \bar{r}^3+38 \bar{r}^4\right)}{12 \left(6 M^2-5 M \bar{r}+\bar{r}^2\right)^2}, \quad
a_{360}=\frac{\bar{r}^3 \left(-66 M^3+69 M^2 \bar{r}-43 M \bar{r}^2+14 \bar{r}^3\right)}{48 (3 M-\bar{r}) (\bar{r}-2 M)^2}, \nonumber \\
a_{504}=\frac{M \bar{r}^2 \left(195 M^3-207 M^2 \bar{r}+83 M \bar{r}^2-9 \bar{r}^3\right)}{(3 M-\bar{r})^3 (\bar{r}-2 M)^2}, \quad
a_{522}=\frac{\bar{r}^2 \left(-279 M^4+307 M^3 \bar{r}-86 M^2 \bar{r}^2-12 M \bar{r}^3+4 \bar{r}^4\right)}{4 (\bar{r}-3 M)^2 (\bar{r}-2 M)^3}, \nonumber \\
a_{540}=\frac{\bar{r}^2 \left(132 M^4-75 M^3 \bar{r}-41 M^2 \bar{r}^2+32 M \bar{r}^3-2 \bar{r}^4\right)}{48 (3 M-\bar{r}) (\bar{r}-2 M)^4}, \quad
a_{702}=-\frac{M \bar{r} \left(89 M^3-73 M^2 \bar{r}+10 M \bar{r}^2+4 \bar{r}^3\right)}{4 (\bar{r}-3 M)^2 (\bar{r}-2 M)^4}, \nonumber \\
a_{720}=\frac{M \bar{r} \left(-19 M^3+7 M^2 \bar{r}+6 M \bar{r}^2-4 \bar{r}^3\right)}{16 (2 M-\bar{r})^5 (3 M-\bar{r})}, \quad
a_{900}=\frac{M^2 \left(2 M^2-2 M \bar{r}+\bar{r}^2\right)}{8 (3 M-\bar{r}) (\bar{r}-2 M)^6}
\end{gather}
and
\begin{gather}
b_{002} = \frac{4 \bar{r}^2 (\bar{r}-2 M)}{\bar{r}-3 M}, \quad
b_{020}=\bar{r}^2, \quad
b_{200}=\frac{\bar{r}}{\bar{r}-2M}
\end{gather}

Next, we compute the effective source corresponding to this singular field. The wave operator in Schwarzschild coordinates is given by
\begin{equation} \label{eq:box-schw}
\Box_{\rm Schw} = -\left(\frac{r}{r-2M}\right) \frac{\partial^2}{\partial t^2} + \left(\frac{r-2M}{r}\right)\frac{\partial^2}{\partial r^2}  + \frac{2(r-M)}{r^2}\frac{\partial}{\partial r}+ \frac{1}{r^2} \frac{\partial^2}{\partial \theta^2} + \frac{1}{r^2 \tan(\theta)}\frac{\partial}{\partial \theta} + \frac{1}{r^2 \sin^2(\theta)} \frac{\partial}{\partial \phi}.
\end{equation}
Applying this to \eqref{eq:PhiS-Schw}, we obtain an effective source of the form
\begin{equation}
S_{\rm eff}^{(4)} = \frac{f(\Delta r, \Delta \theta, Q)}{\left(\sum_{i,j,k = 0}^{i+j+k\le2} b_{ijk} \Delta r^i \Delta \theta^j \Delta Q^k\right)^{11/2}},
\end{equation}
where $f(\Delta r, \Delta \theta, Q)$ is a polynomial in $\Delta r$, $\Delta \theta$, $Q$ (and contains terms involving $\tan\Delta \theta$ and $\sec\Delta \theta$) and the $b_{ijk}$ are the same as those in the singular field.

\begin{figure}
\includegraphics[width=4cm]{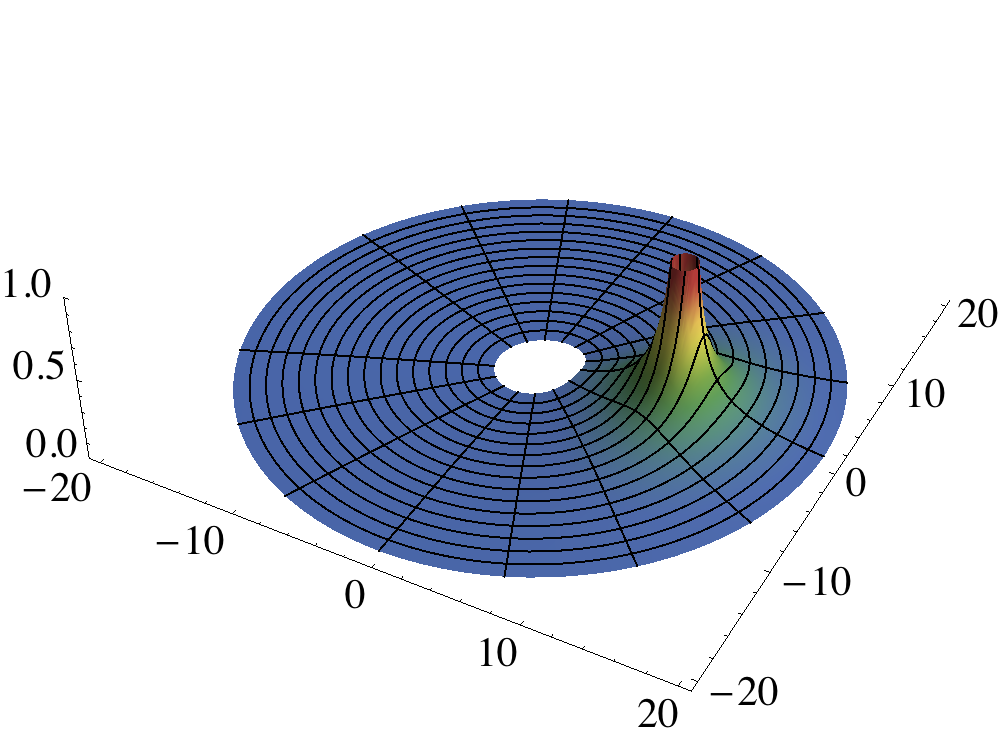}
\includegraphics[width=4cm]{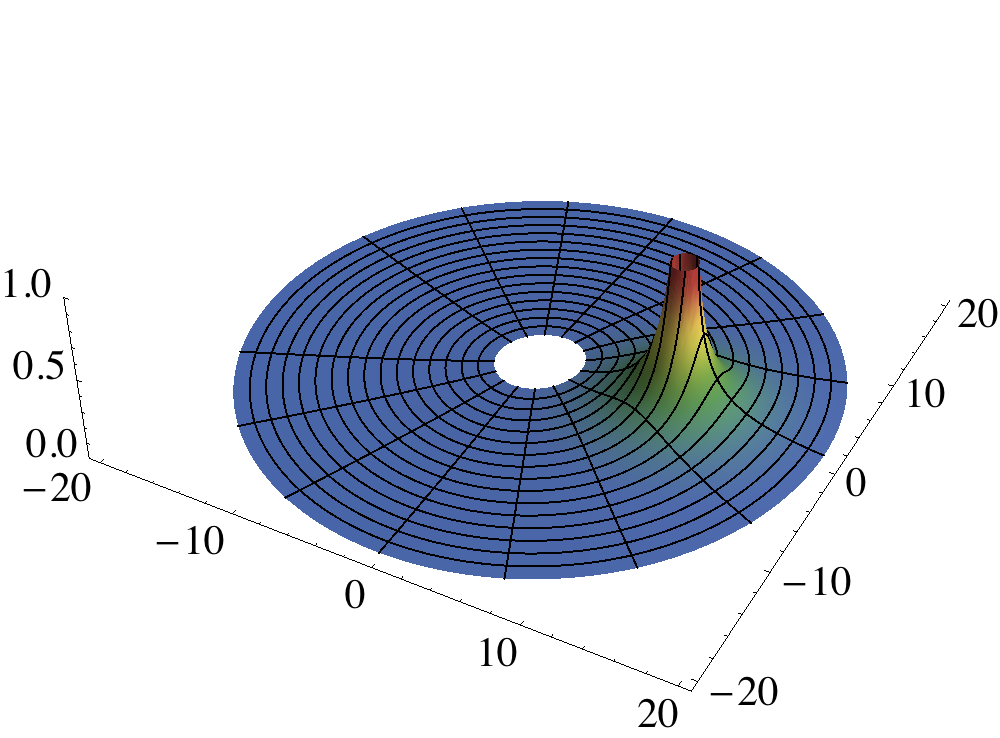}
\includegraphics[width=4cm]{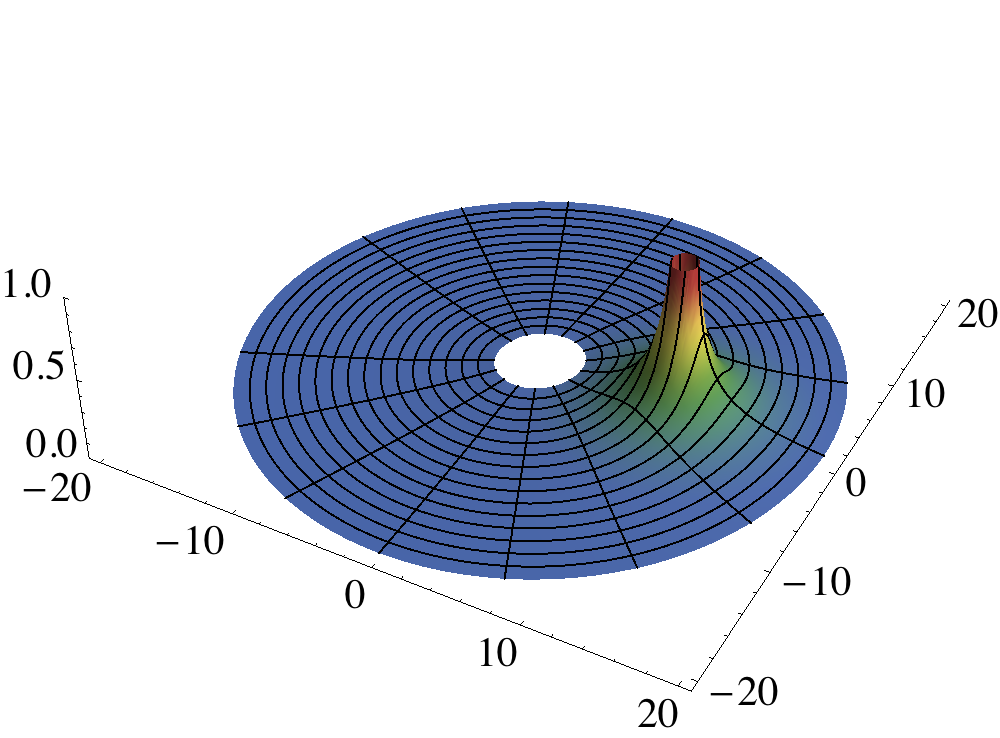}
\includegraphics[width=4cm]{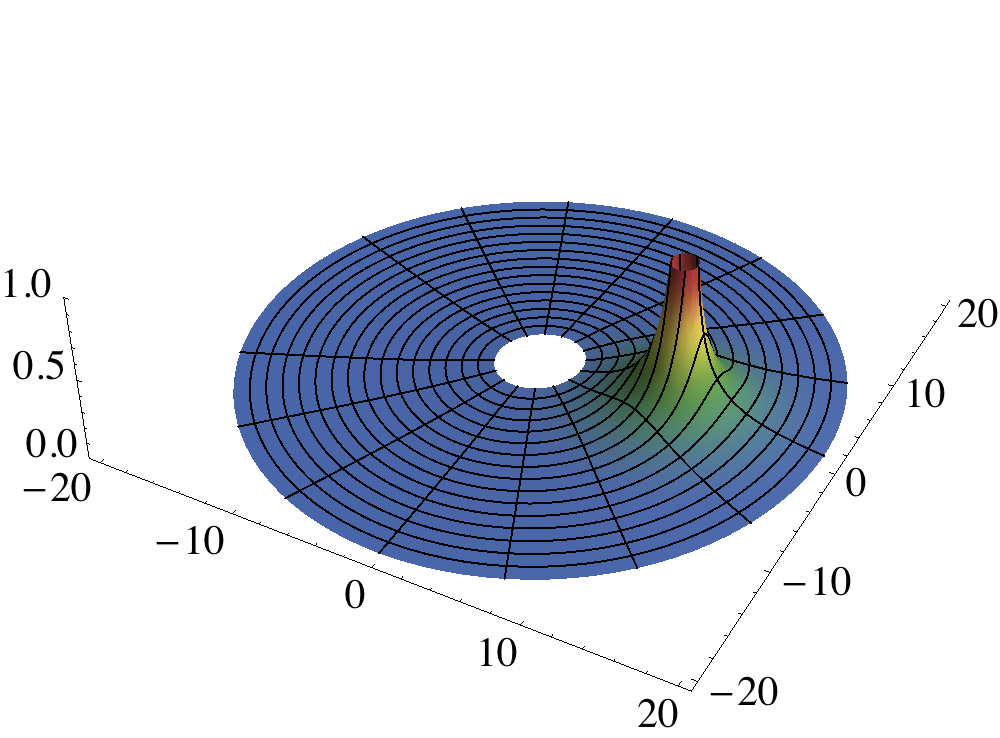}
\includegraphics[width=4cm]{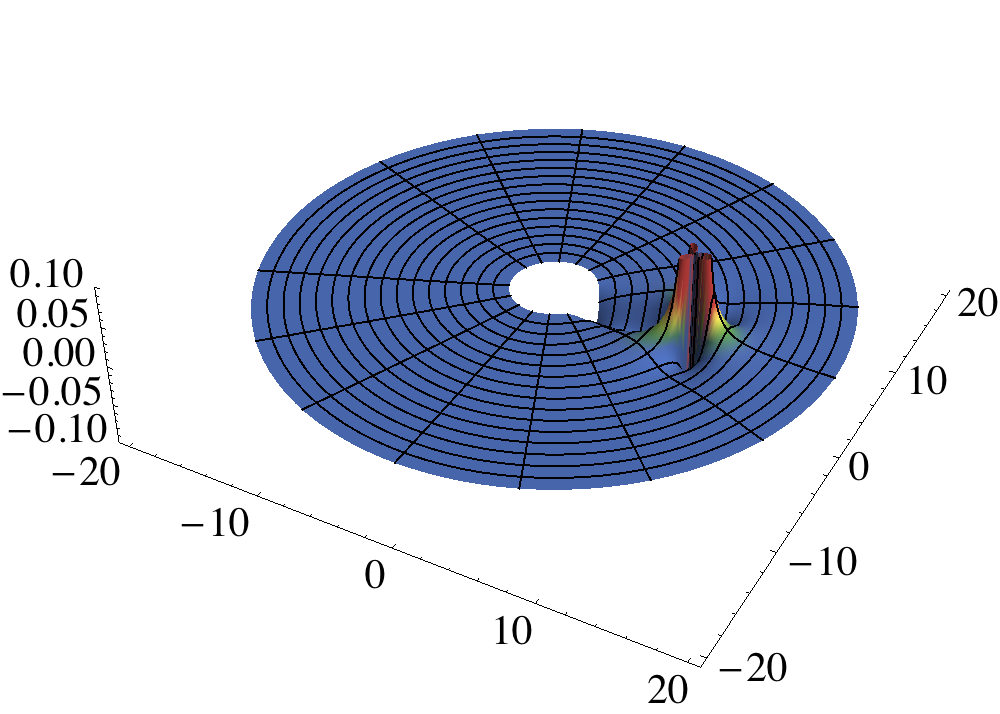}
\includegraphics[width=4cm]{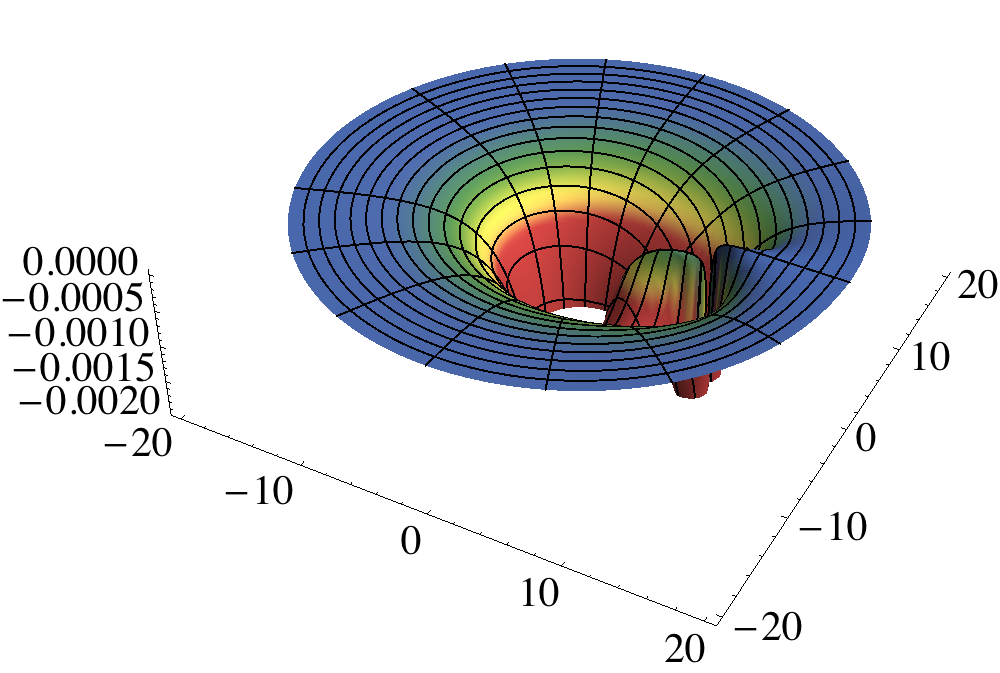}
\includegraphics[width=4cm]{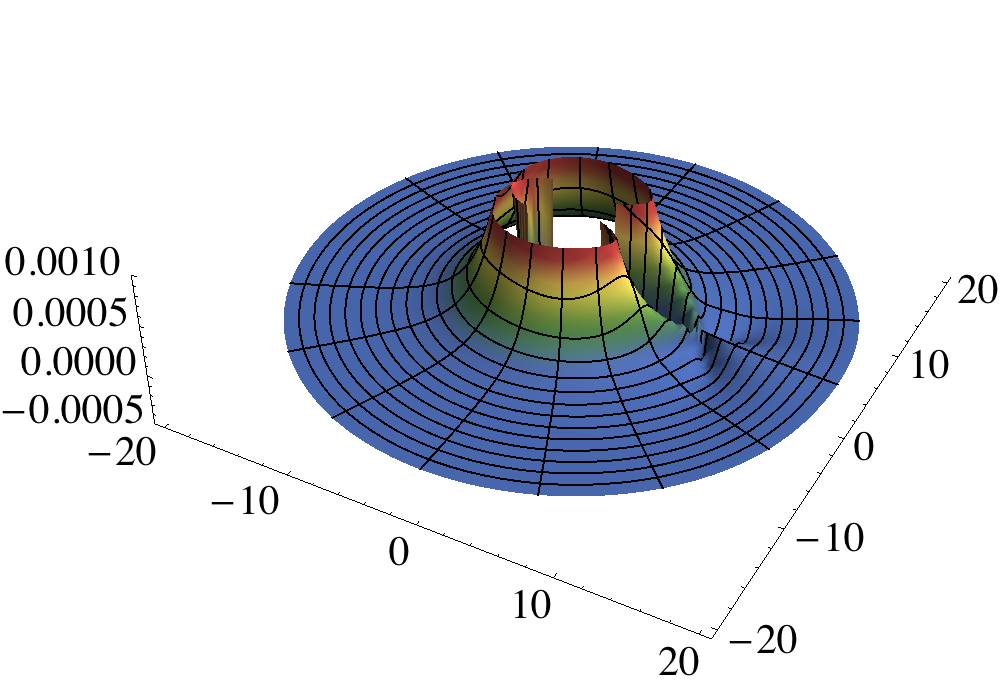}
\includegraphics[width=4cm]{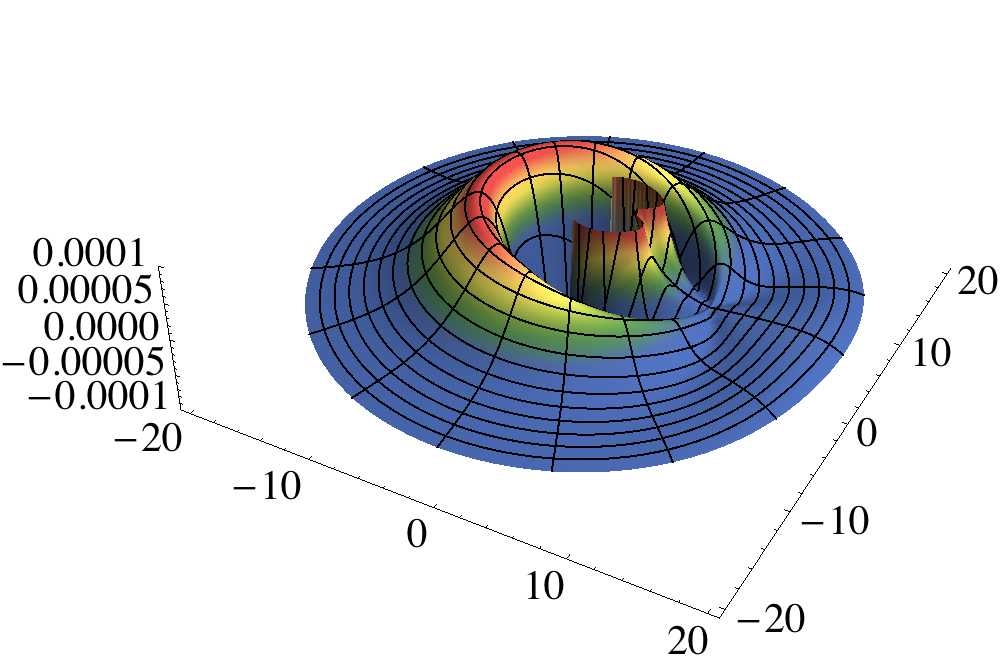}
\caption{Singular field (top) and effective source (bottom) along the equatorial plane for a particle in a circular orbit around a Schwarzschild black hole. From left to right: first, second, third and fourth order cases are shown. Note that in the third and fourth order cases, we used the method described in Sec.~\ref{sec:periodicity} to ensure periodicity in $\phi$.}
\label{fig:schw-results}
\end{figure}

\begin{figure}
\includegraphics[width=4cm]{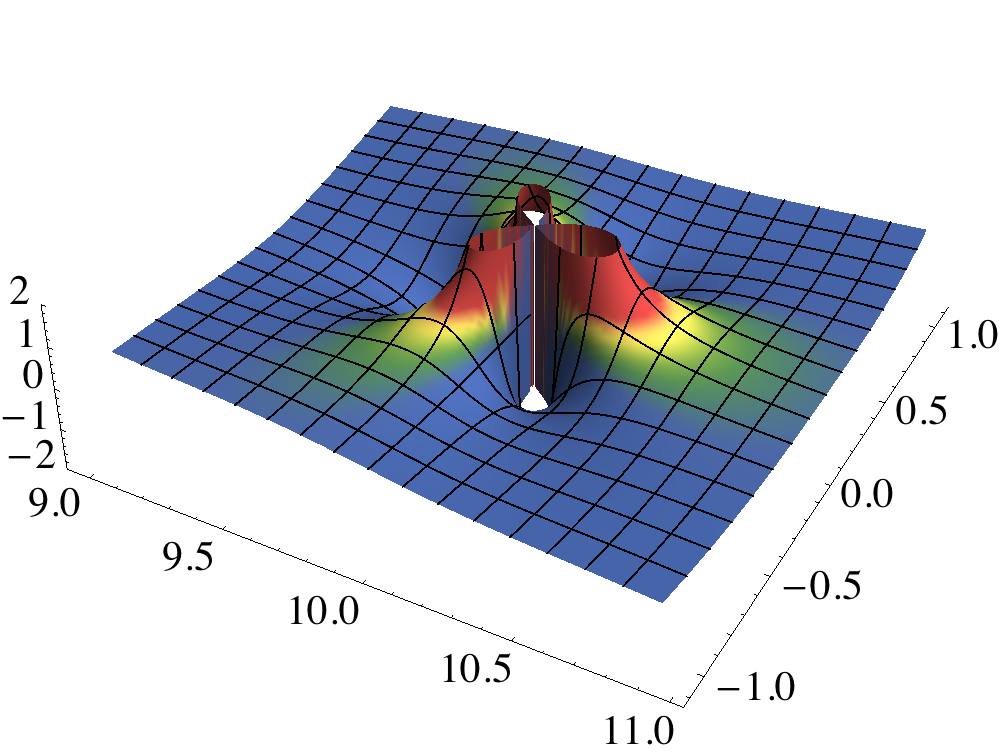}
\includegraphics[width=4cm]{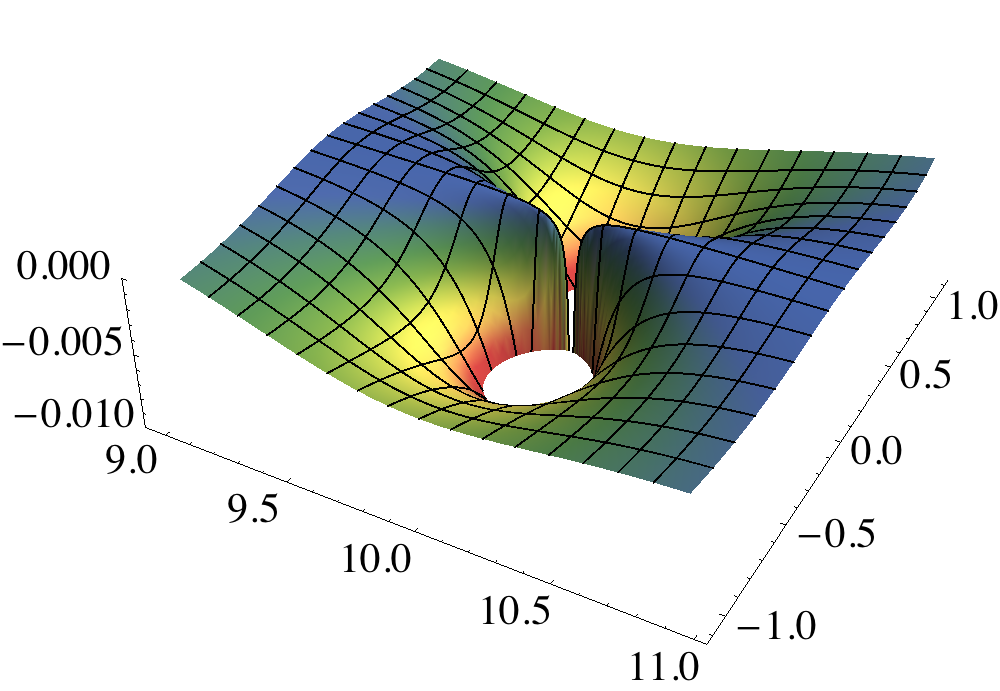}
\includegraphics[width=4cm]{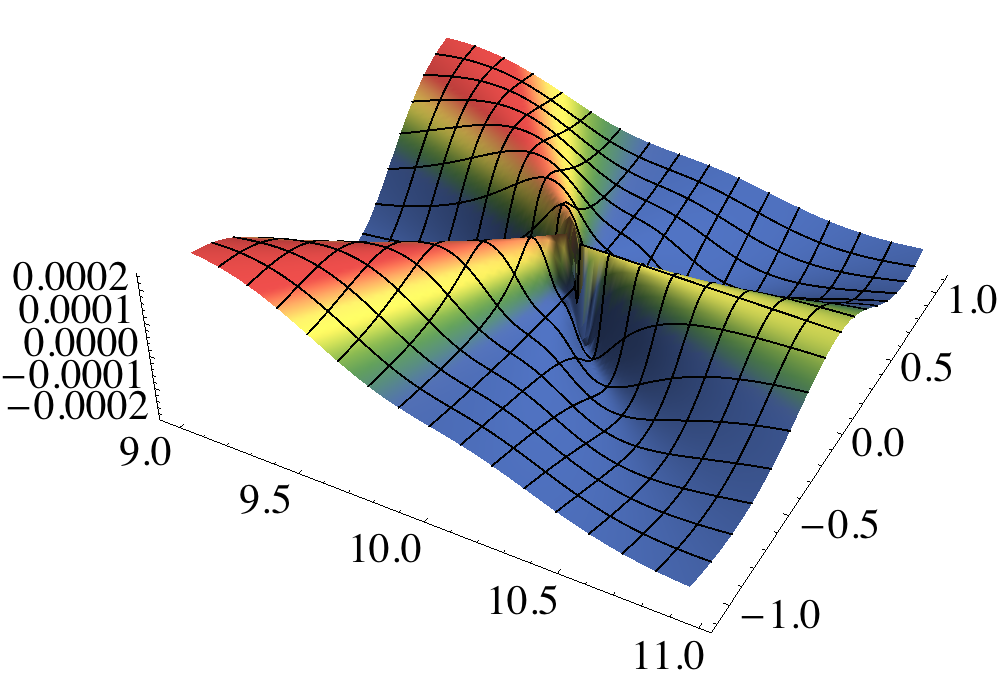}
\includegraphics[width=4cm]{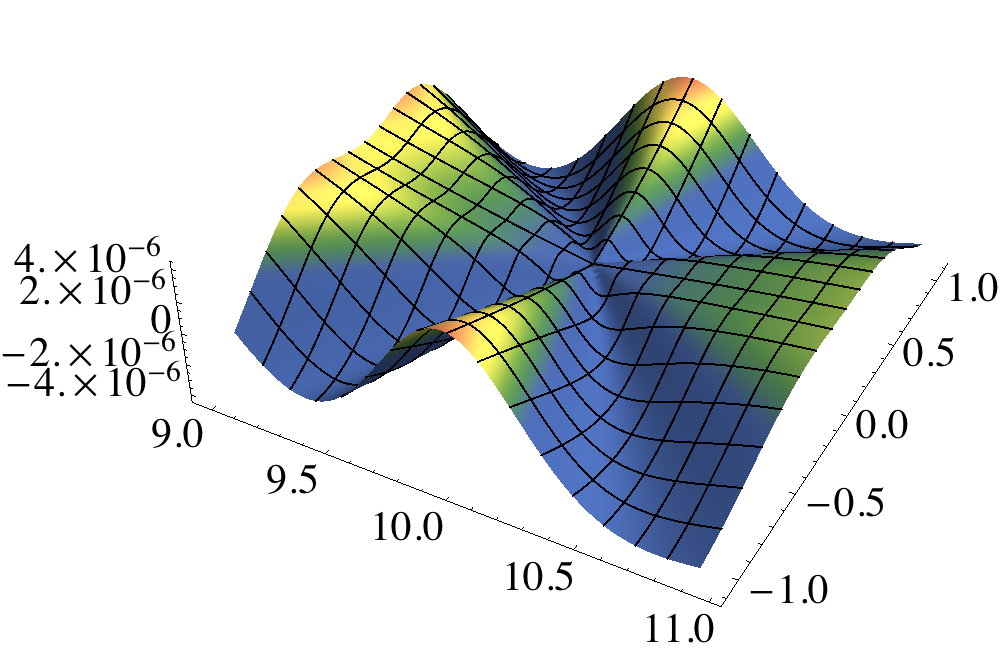}
\includegraphics[width=3.95cm]{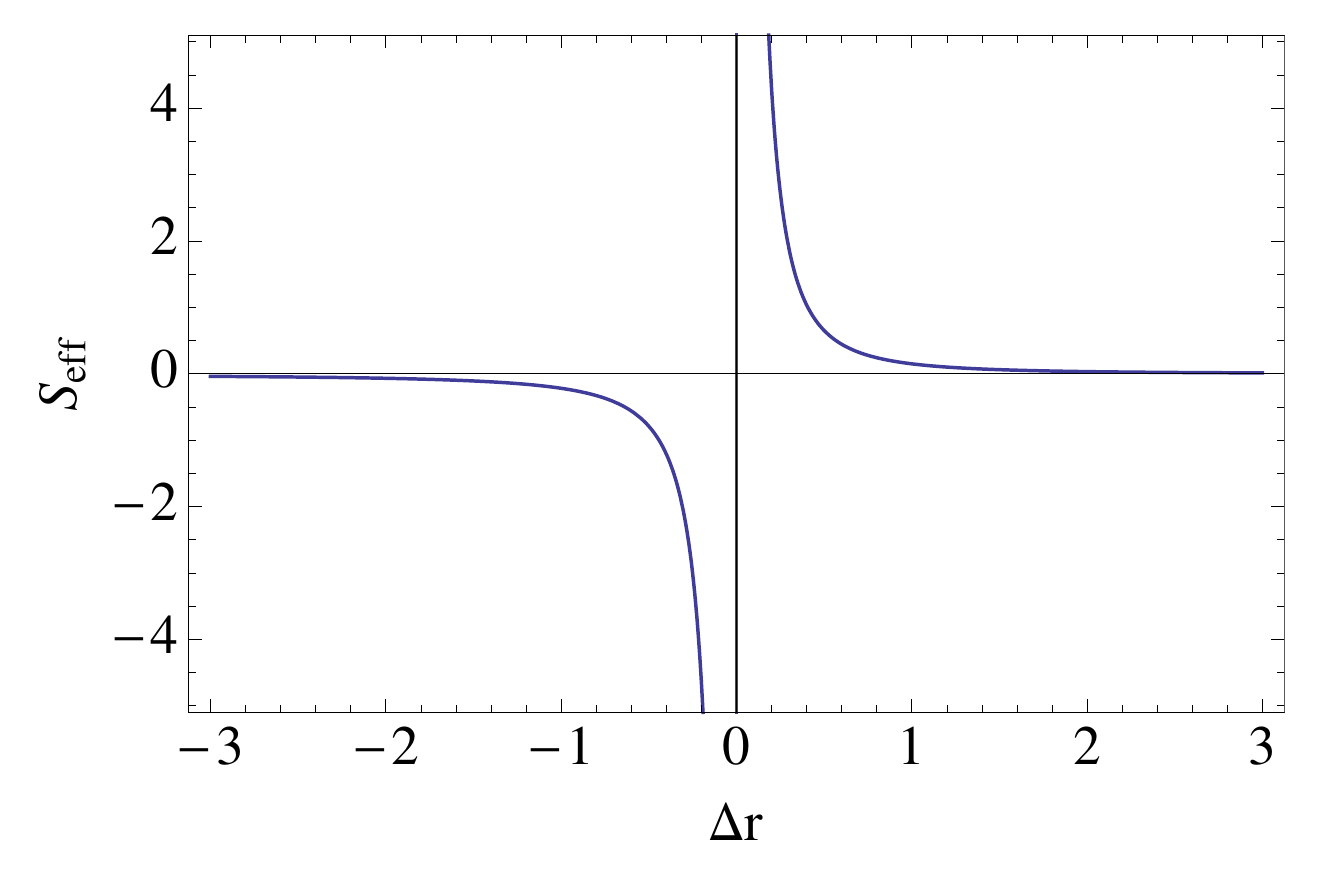}
\includegraphics[width=4.25cm]{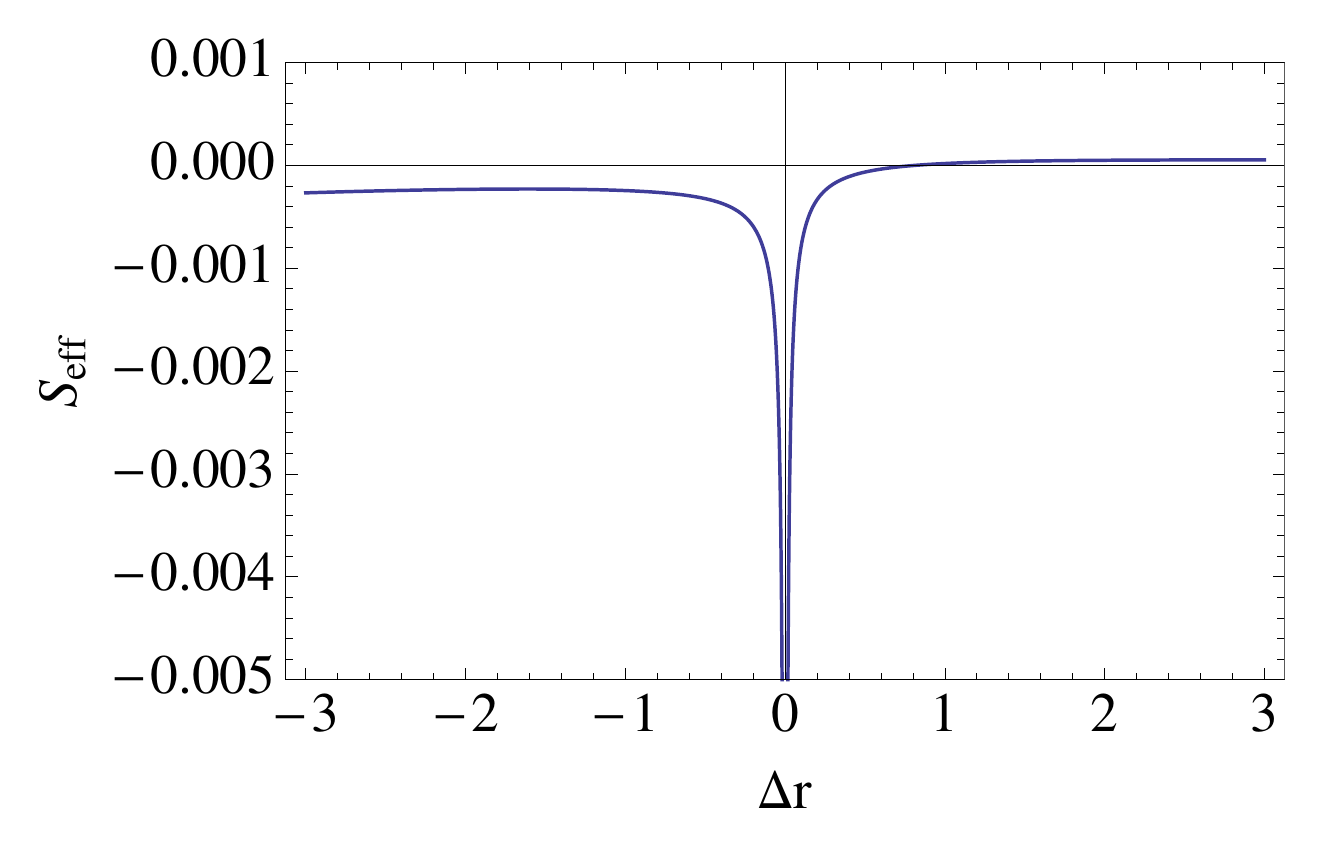}
\includegraphics[width=4.5cm]{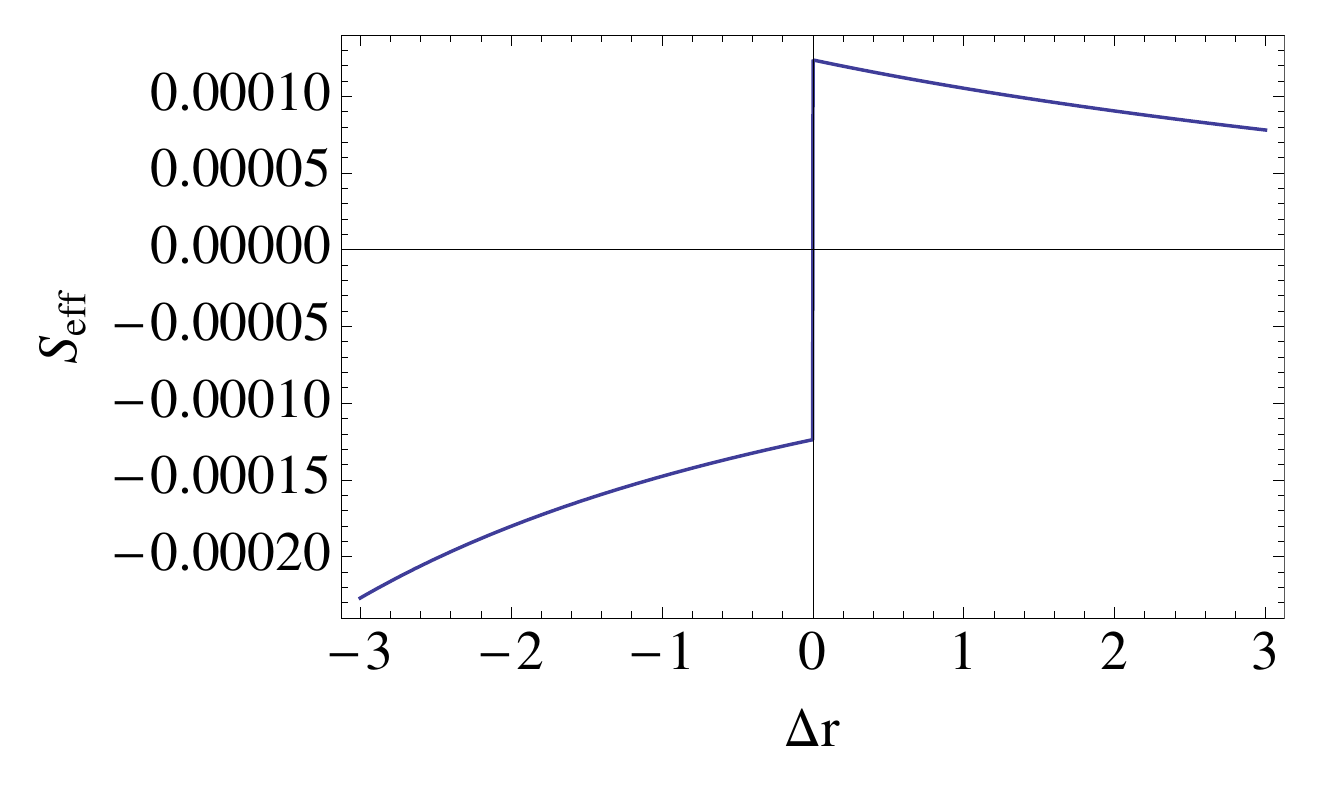}
\includegraphics[width=4.5cm]{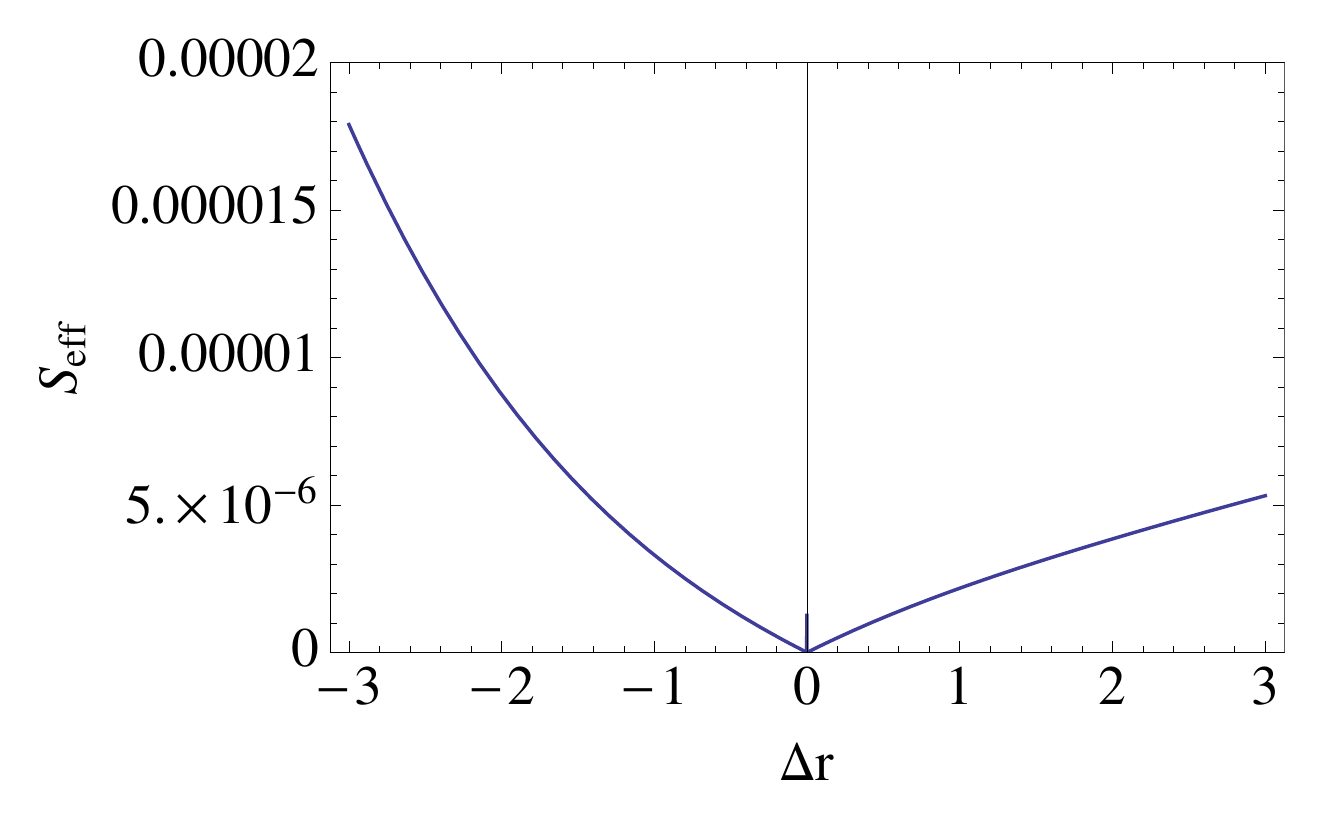}
\caption{Close-up view of the effective source along the equatorial plane (top) and along a radial slice through the particle (bottom) for a particle in a $r=10M$ circular orbit around a Schwarzschild black hole. From left to right: first, second, third and fourth order cases are shown.}
\label{fig:schw-results2}
\end{figure}

In Fig.~\ref{fig:schw-results} we plot the first, second, third and fourth order singular field and corresponding effective source for the case of a particle in a circular orbit at $\bar{r} = 10M$ in Schwarzschild. All four cases have a visually similar singular field. This is not surprising given they share the same singular behaviour and only differ in higher order corrections. The corresponding effective source, however, is very different. As expected from the discussion of Sec.~\ref{sec:effsrc-approx}, at first and second order the effective source diverges at the particle, while at third and fourth order it is finite. Figure \ref{fig:schw-results2} shows a zoomed in view of the effective source in each case, along with a slice along the radial direction, passing through the particle. From this we see more clearly the behaviour of the effective source near the particle: at first order it is $C^{-3}$, at second order it is $C^{-2}$, at third order it is $C^{-1}$ and at fourth order it is $C^0$.

\subsection{Circular geodesic in Kerr spacetime}
\label{sec:Kerr-circular}
To compute the singular field and effective source in Kerr spacetime, we consider its metric in Boyer-Lindquist coordinates,
\begin{equation}
\label{eq:KerrMetric}
ds^2 = - \left(1-\frac{2Mr}{\Sigma}\right)dt^2 
			- \frac{4aMr\sin^2\theta}{\Sigma}dtd\phi
			+ \frac{\Sigma}{\Delta}dr^2\\
			+ \Sigma d\theta^2
			+ \left(\Delta+\frac{2Mr(r^2+a^2)} {\Sigma}\right) \sin^2\theta d\phi^2
\end{equation}
where
\begin{equation}
\Sigma = r^2+a^2\cos^2\theta, \qquad
\Delta = r^2-2Mr+a^2.
\end{equation}
As in the Schwarzschild case, in order to obtain sufficiently compact expressions to be given here, we assume that the motion follows a circular, prograde equatorial geodesic, i.e.\ \cite{Chandrasekhar}
\begin{gather}
\bar{r} = \text{constant}, \quad \bar{\theta} = \frac{\pi}{2}\nonumber \\
 u^r = 0, \quad u^\theta = 0, \quad u^\phi = \frac{\sqrt{Mr}}{r\sqrt{r^2-3Mr+2a\sqrt{Mr}}}, \quad u^t = \frac{aM+\sqrt{Mr^3}}{\sqrt{Mr}\sqrt{(r^2-3Mr+2a\sqrt{Mr})}}.
\end{gather}
We also use the freedom in the choice of $t$ to set the field point and world-line point to be at the same coordinate time, i.e.
\begin{equation}
\bar{t} = t \qquad \bar{\phi} = \Omega t
\end{equation}
where
\begin{equation}
\Omega = \frac{M}{aM+\sqrt{M\bar{r}^3}}
\end{equation}
is the orbital frequency. Combining everything, we obtain a fourth order approximation to the singular field of a scalar charge on a circular equatorial orbit around a Kerr black hole:
\begin{equation}
\label{eq:PhiS-Kerr}
\tilde{\Phi}_{\rm S}^{(4)}(r, \theta, \phi, t) = \frac{\sum_{i,j,k = 0}^{i+j+k\le9} a_{ijk} \Delta r^i \Delta \theta^j Q^k}{\left(\sum_{i,j,k = 0}^{i+j+k\le2} b_{ijk} \Delta r^i \Delta \theta^j Q^k\right)^{7/2}}
\end{equation}
where $\Delta r = r-\bar{r}$, $\Delta \theta = \theta - \pi/2$, $Q = \sin\big(\frac12(\phi - \Omega t)\big)$ and 
where the non-zero coefficients, $a_{ijk}$ and $b_{ijk}$ are functions of the orbital radius, $\bar{r}$, and the spin parameter, $a$, and are given by taking the expressions in Ref.~\cite{dolan-etal:11}, making the change of variables $\Delta \phi \to Q$ and re-expanding as
described in Sec.~\ref{sec:PhiS-practical}.

Next, we compute the effective source corresponding to this singular field. The wave operator in Kerr (Boyer-Lindquist) coordinates is given by
\begin{IEEEeqnarray}{rCl} \label{eq:box-kerr}
\Box_{\rm BL} &=& 
-\left[1+\frac{2M r\left(a^2+r^2\right)}{\left(a^2+r^2-2Mr\right) \left(r^2+a^2 \cos^2 \theta \right)}\right]\frac{\partial^2}{\partial t^2} 
 +\frac{ \left(a^2+r^2-2Mr\right)}{r^2+a^2 \cos^2 \theta} \frac{\partial^2}{\partial r^2}
 + \frac{2 (r-M)}{r^2 + a^2 \cos^2 \theta}\frac{\partial}{\partial r} \nonumber \\
 &&
 +\frac{1}{r^2+a^2 \cos^2\theta}\frac{\partial^2}{\partial \theta^2}
 +\frac{\cot \theta }{r^2+a^2 \cos^2\theta}\frac{\partial}{\partial \theta}
 +\frac{ \left(r^2-2Mr+a^2 \cos^2 \theta \right)\csc^2 \theta }{\left(a^2+r^2-2Mr\right) \left(r^2+a^2 \cos^2 \theta \right)}\frac{\partial^2}{\partial \phi^2} \nonumber \\
 &&
 -\frac{4 a M  r}{\left(a^2+r^2-2Mr\right) \left(r^2+a^2 \cos^2 \theta \right)}\frac{\partial^2}{\partial \phi \partial t}.
\end{IEEEeqnarray}
Applying this to \eqref{eq:PhiS-Kerr}, we obtain an effective source of the form
\begin{equation}
S_{\rm eff}^{(4)} = \frac{f(\Delta r, \Delta \theta, Q)}{\left(\sum_{i,j,k = 0}^{i+j+k\le2} b_{ijk} \Delta r^i \Delta \theta^j Q^k\right)^{11/2}},
\end{equation}
where $f(\Delta r, \Delta \theta, Q)$ is a polynomial in $\Delta r$, $\Delta \theta$, $Q$ (and contains terms involving $\tan \Delta \theta$ and $\sec\Delta \theta$) and the $b_{ijk}$ are the same as those in the singular field.

\begin{figure}
\includegraphics[width=4cm]{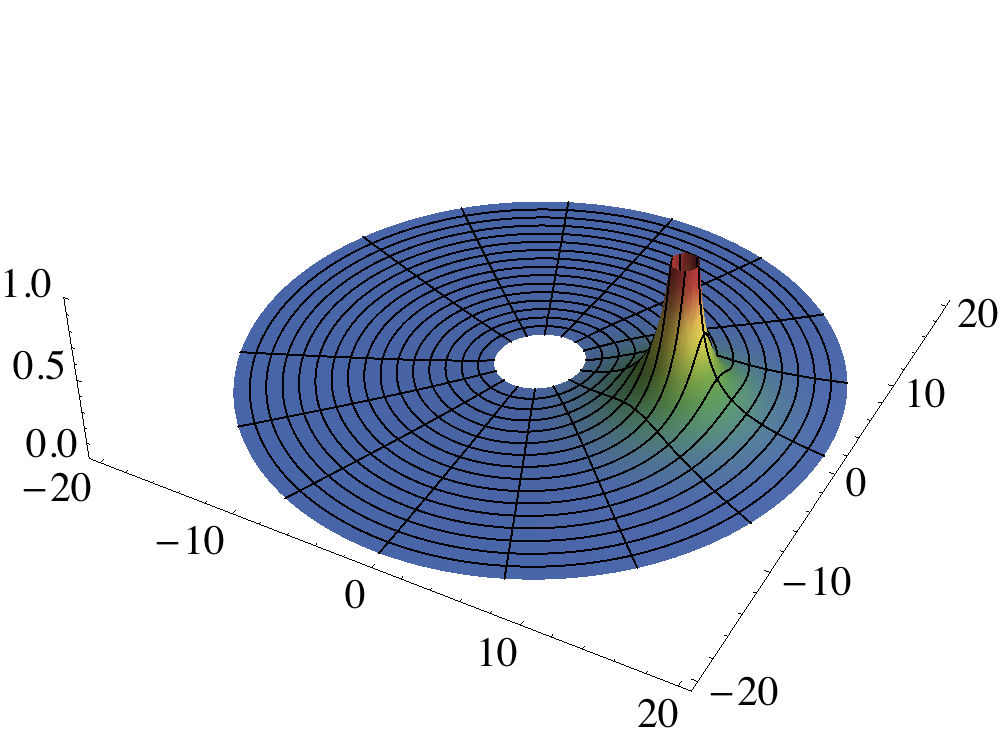}
\includegraphics[width=4cm]{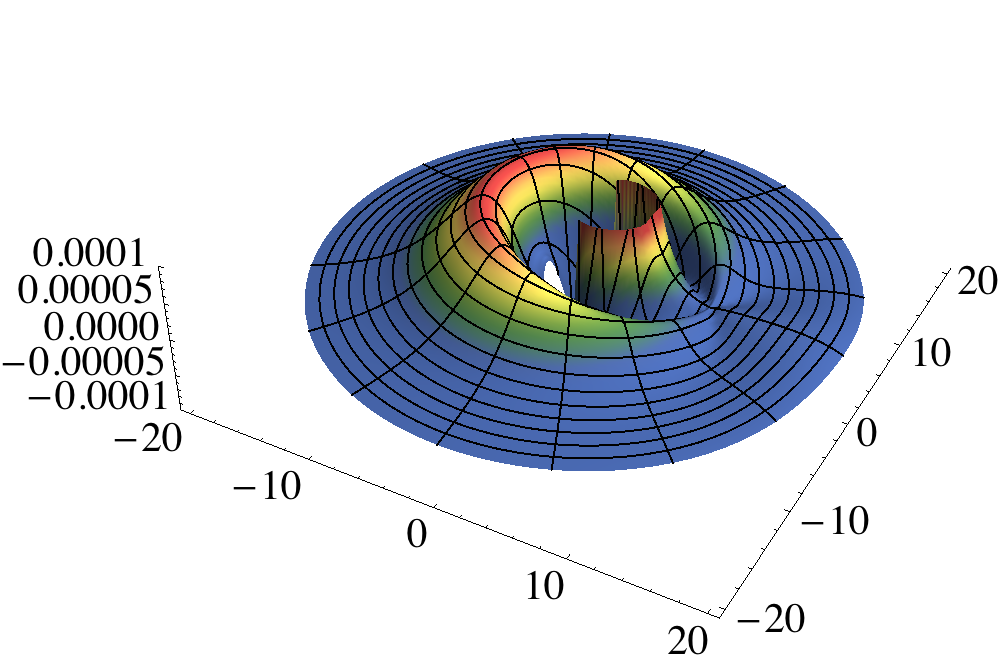}
\includegraphics[width=4cm]{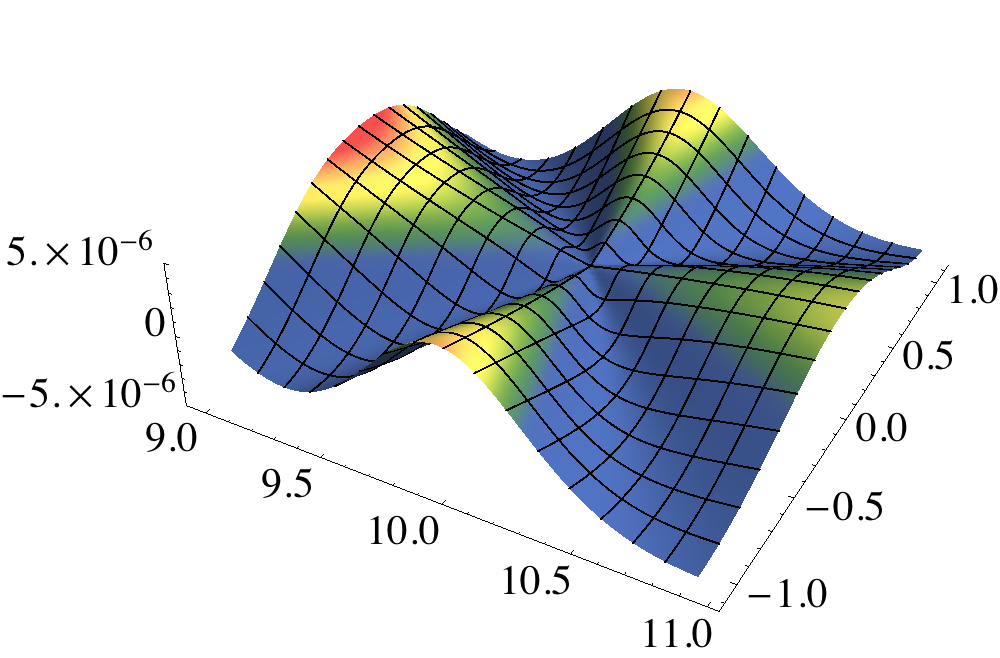}
\includegraphics[width=4cm]{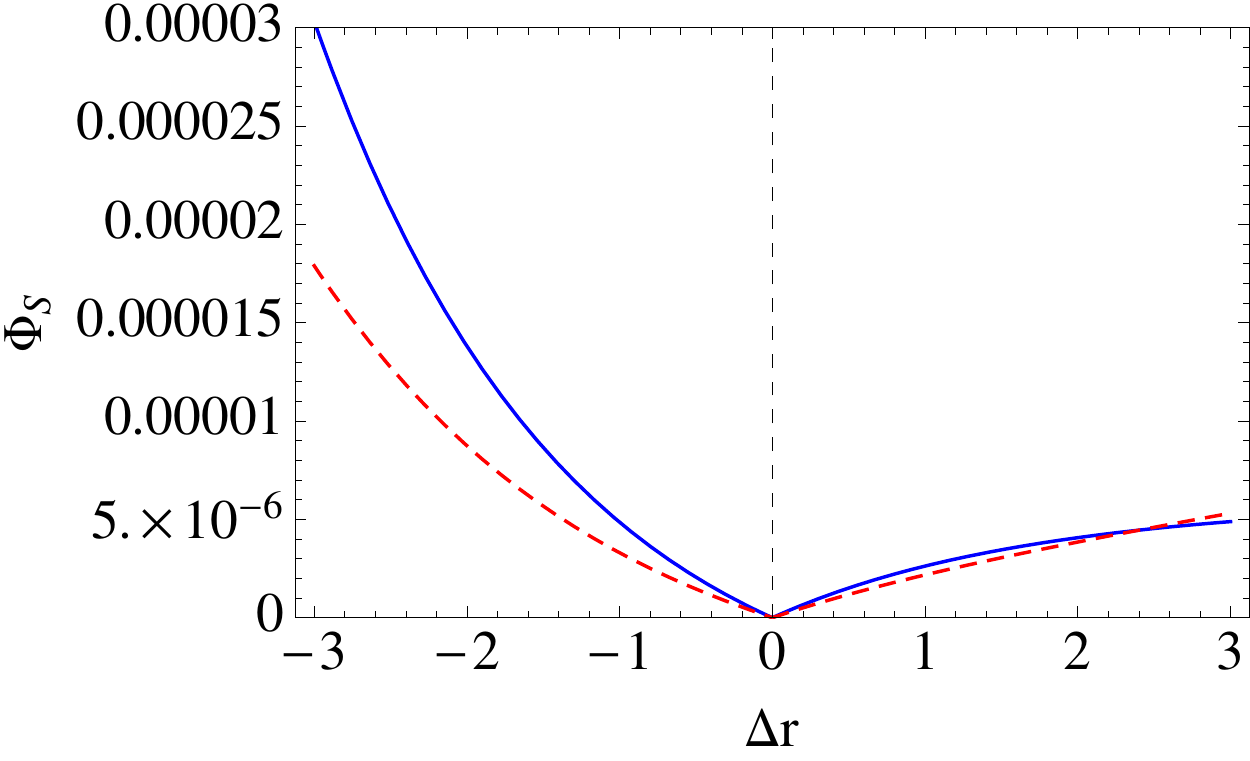}
\caption{Particle following a circular equatorial geodesic around a Kerr black hole with spin $a=0.99M$. Left to right: (1) fourth order singular field, (2) fourth order effective source, (3) close-up view of the effective source and (4) effective source along a radial slice through the particle.
In (4), we also show the Schwarzschild result as a dashed red line for comparison.
Note that we used the method described in Sec.~\ref{sec:periodicity} to ensure periodicity in $\phi$.}
\label{fig:kerr-results}
\end{figure}

In Fig.~\ref{fig:kerr-results} we plot the fourth order singular field and corresponding effective source for the case of a particle in a circular orbit at $\bar{r} = 10M$ around a Kerr black hole with spin $a=0.99M$.  As expected from the discussion of Sec.~\ref{sec:effsrc-approx}, the fourth order effective source is finite and continuous, i.e.\ $C^0$.
In the rightmost figure, we compare against the equivalent case in Schwarzschild. Both cases are qualitatively remarkably similar, only differing significantly in magnitude close to the black hole.

\subsection{Generic geodesic in Kerr spacetime}
\label{sec:generic-kerr}
To illustrate the power of the method developed here, we now consider a more generic configuration. We choose an arbitrary timelike geodesic of the Kerr spacetime and compute the singular field and effective source at a point along that geodesic. In particular, we make the choice
\begin{gather}
\bar{r} = 10M, \quad \bar{\theta}=\frac{\pi}{2}, \quad a = 0.99M, \quad M=1 \nonumber \\
u^\phi = \frac{\sqrt{Mr}}{r\sqrt{r^2-3Mr+2a\sqrt{Mr}}}, \quad u^\theta = \frac{1}{2} u^\phi, \quad u^r = M u^\phi
\end{gather}
with $u^t$ being determined by the normalization of the four-velocity, $u_\alpha u^\alpha = -1$. The computation of the singular field and effective source proceeds exactly as in the circular orbit case, the only difference being that the resulting expressions are larger. In fact, they are too large to be useful in printed form. Since they are relatively manageable with computer algebra, however, we have made them available as \emph{Mathematica} code \cite{EffSource-online}.

\begin{figure}
\includegraphics[width=4cm]{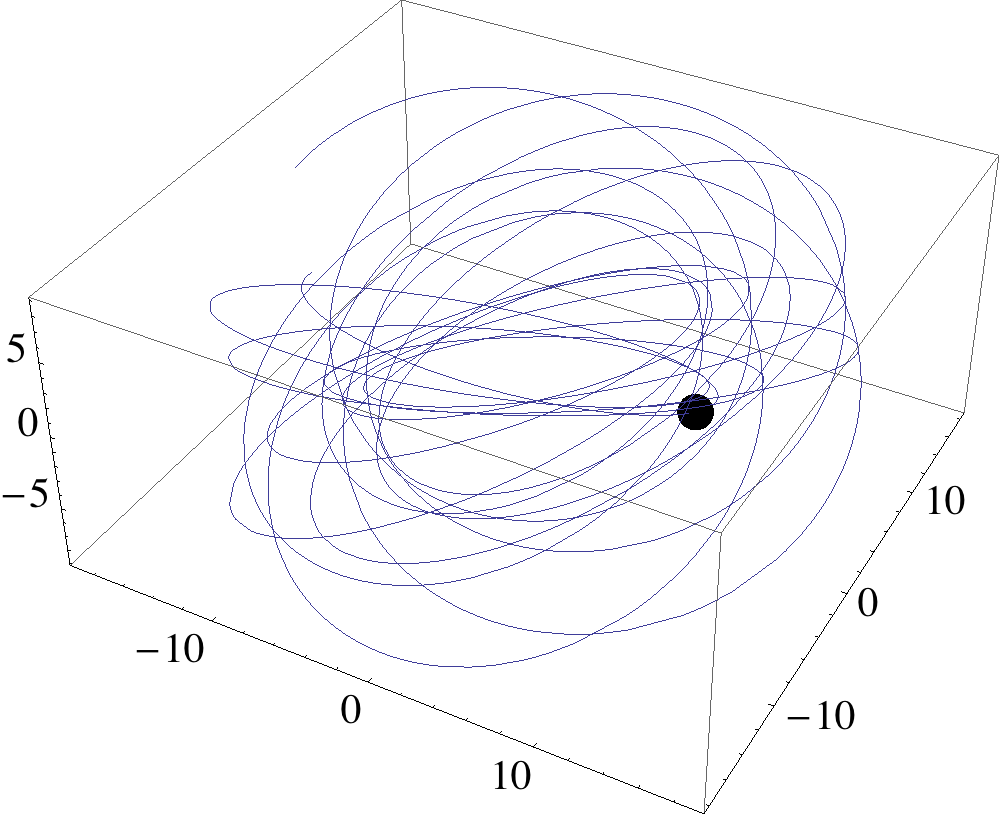}
\includegraphics[width=4cm]{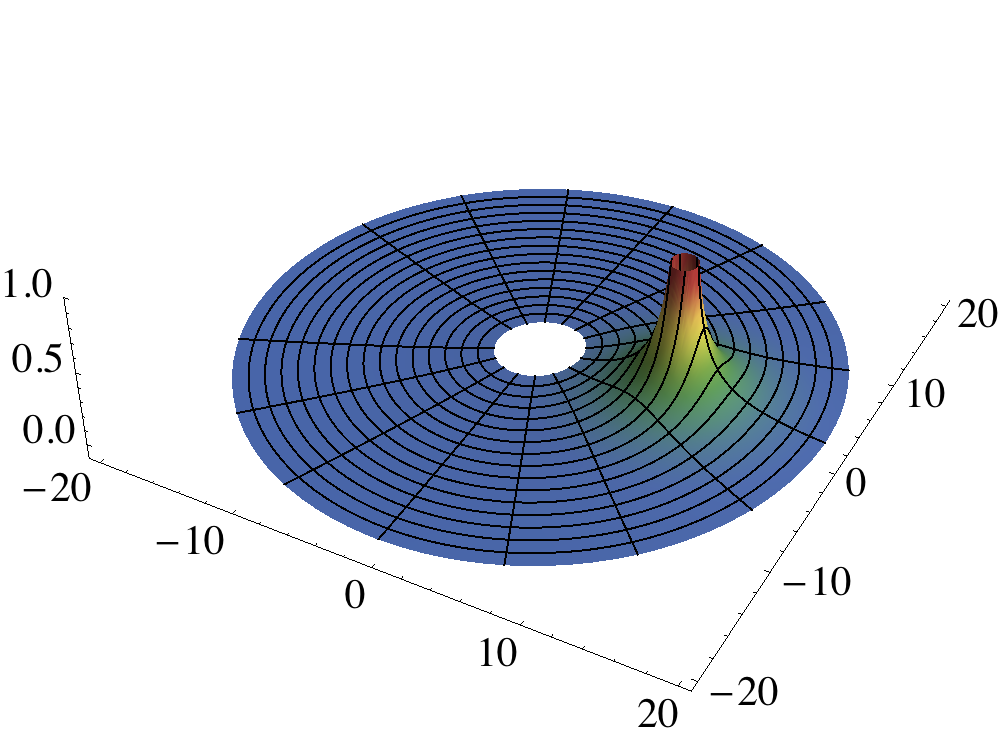}
\includegraphics[width=4cm]{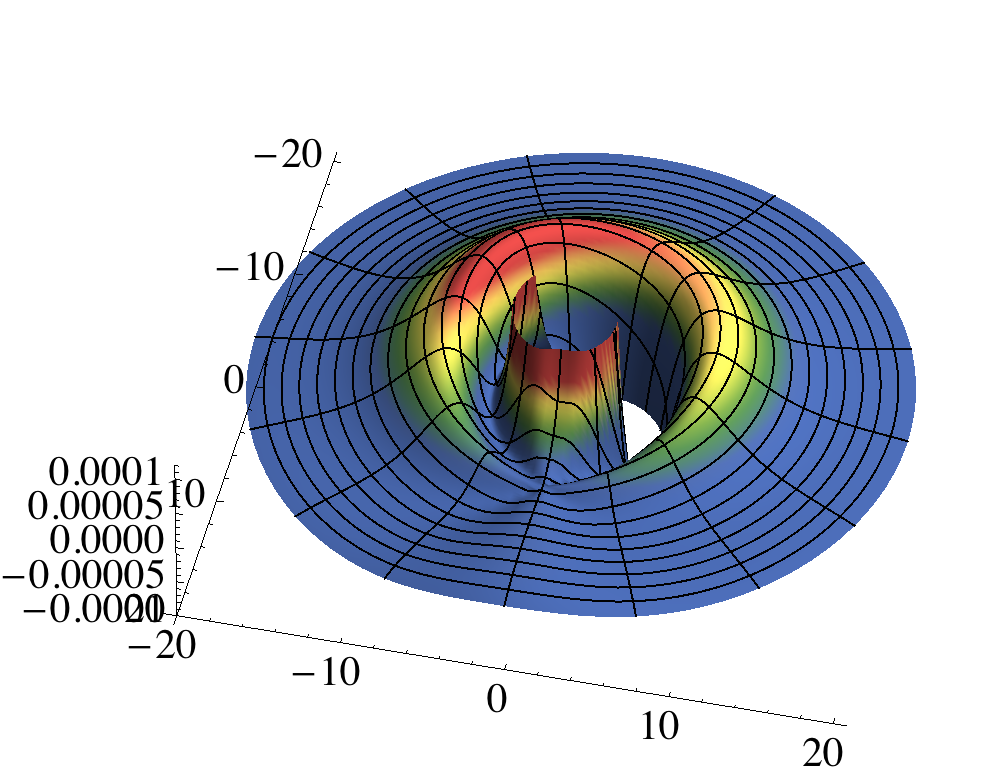}
\includegraphics[width=4cm]{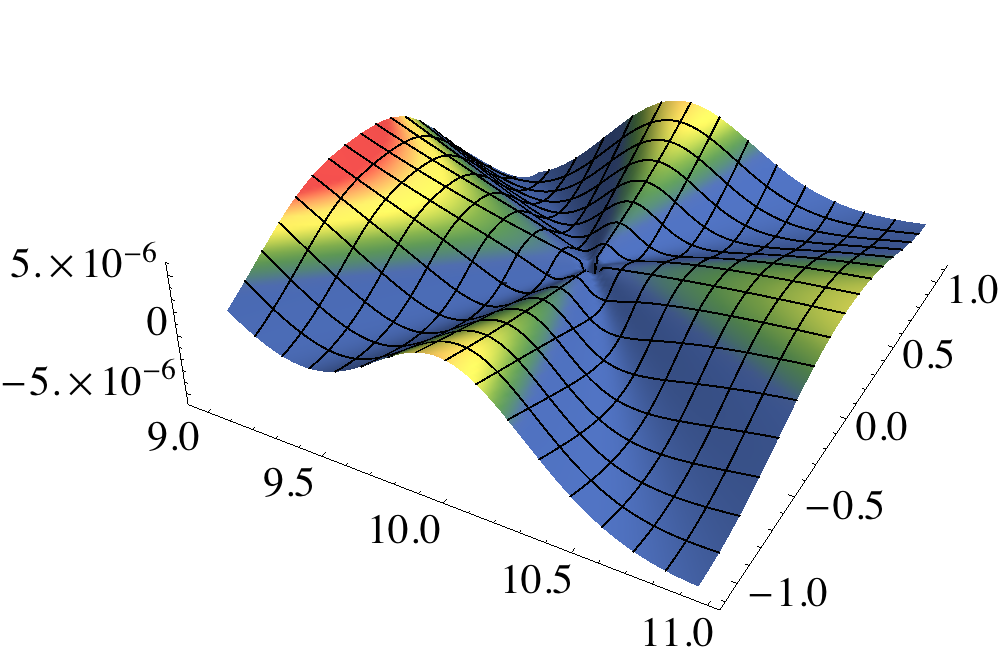}
\caption{Particle following generic geodesic around a Kerr black hole with spin $a=0.99M$. Left to right: (1) particle's world-line (solid line) with position at which the singular field and effective source are computed indicated by a black dot, (2) fourth order singular field, (3) fourth order effective source, (4) close-up view of the effective source. Note that we used the method described in Sec.~\ref{sec:periodicity} to ensure periodicity in $\phi$.}
\label{fig:kerr-generic-results}
\end{figure}

In Fig.~\ref{fig:kerr-generic-results}, we illustrate the behaviour of the singular field and effective source for this configuration. The leftmost plot shows the geodesic over several orbits, indicating that it is both inclined and eccentric. The black dot on this plot indicates the point $\bar{r}=10M$, $\bar{\theta}=\pi/2$, $\bar{\phi}=0$ at which the singular field and effective source in the subsequent plots is computed. The second plot shows the singular field along the equatorial plane. The third and fourth plots show the effective source along the equatorial plane. As expected, this fourth order effective source is continuous, but not differentiable at the particle.

\section{Discussion and summary} 
\label{sec:discussion}

In this paper, we have developed an approximation to the singular field of a point scalar charge to quadratic order in the distance from the charge.  This is sufficient to give second order convergence in the grid spacing for $3+1$D numerical calculations and to give $m^{-4}$ and $l^{-4}$ convergence in the $m$-mode and $l$,$m$-mode schemes, respectively. To go to higher order (for better convergence) one would need to:
\begin{enumerate}
\item Calculate the higher order terms in the coordinate expansions of $\sigma_{\bar{a}}$. This is a recursive calculation and the expressions get more unwieldy as the order increases. However, the calculation method is general and only limited by computational power.

\item Calculate higher order corrections to $\sigma_{\alpha'} u^{\alpha'}$ and $\sigma_{\alpha''} u^{\alpha
''}$. This is straightforward using higher order covariant expansions of $\sigma$ and its derivatives \cite{Haas:Poisson:2006}, which are easily obtained to much higher order than is needed here using non-recursive methods \cite{Transport}.

\item Calculate higher order terms in the series expansions of $U(x,x')$, $U(x,x'')$ and $V(x,z(\tau))$. These are also easily obtained from the semi-recursive methods of Ref.~\cite{Transport}.
\end{enumerate}
The calculation of a higher order singular field and effective source is therefore a straightforward (if somewhat tedious) process. With the expressions becoming more unwieldy at each order, one must balance the calculation effort against the benefits of doing so. It seems likely that the fourth order approximation presented here is the `sweet spot', giving reasonably good convergence with modest computational difficulty.

As shown in Sec.~\eqref{sec:generic-kerr}, this method works for very general motion in the Kerr spacetime. Furthermore, although no explicit calculations have been done here for other spacetimes, it is clear that Eq.~\eqref{eq:PhiS-approx} is valid in any spacetime. It would therefore be straightforward to apply this method to any (not necessarily Ricci-flat) spacetime. As the primary motivation of this work has been to study rotating black holes, however, we have chosen to only consider Kerr and Schwarzschild spacetimes in detail in this work.

The methods presented here are useful for computing expressions for the singular field and effective source for generic configurations. The actual evolution of a wave equation with this source, along with the calculation of the self-force should be explored separately. Here, we simply note that we 
have implemented two separate numerical evolution codes using the
singular field and effective source presented here: one uses the
window-function approach with a $3+1$D numerical evolution;
the other uses the world-tube approach followed by an $m$-mode
decomposition and a separate $2+1$D numerical evolution for each~$m$.
We have verified that both codes give correct results
(as determined by comparison with frequency-domain calculations).
Further details of these codes will be presented elsewhere, with some results already having been
published. Using a separate numerical code, Dolan and Barack \cite{Dolan:Barack:2010} evolved the $2+1$D scalar wave equation (with the singular field and effective source as given in Sec.~\ref{sec:Schw-circular}\footnote{In fact, the singular field used by Dolan and Barack differs slightly from that of Sec.~\ref{sec:Schw-circular}. Nonetheless, it was computed using the same methods and differs only at higher order than the order of the approximation.}) for a particle in a circular orbit around a Schwarzschild black hole. This 
calculation was subsequently extended to the case of circular orbits in
Kerr spacetime in \cite{dolan-etal:11,Thornburg:Capra} with further progress toward generic configurations in Kerr spacetime under way.
In a recent work \cite{diener-etal:12a}, the effective source presented here was used to self-consistently evolve the orbit of a point scalar
charge in the Schwarzschild spacetime, incorporating the back reaction from the self-force into
the evolution.

One major issue remaining in the effective source approach is the computational efficiency of the source calculation. For the approach to be of practical use, its calculation must be sufficiently fast that it does not have a prohibitive impact on the run time of a numerical code. This is a serious concern - the expression for the fourth order effective source may be dramatically larger than a finite difference representation of the wave equation, for example. Some steps have been taken in this paper to improve the efficiency of the source calculation. In Appendix \ref{sec:numerical}, we discuss some specific methods for evaluating the effective source as efficiently as possible. As mentioned in Sec.~\ref{sec:PhiS-practical}, we have also made use of specific choices for the singular field in an effort to minimize the size of the resulting expressions. Despite these efforts, the reality is that the calculation of the effective source will considerably affect the run time of a numerical code.

Fortunately, there remain several possibilities for further optimization. The advent of GPU (Graphics Processing Unit) computing has allowed for dramatic performance improvements in certain applications. It seems likely that the embarrassingly parallel nature of the effective source calculation on a grid of points is an ideal candidate for implementation in a GPU programming framework such as CUDA or OpenCL. Given other applications have seen speed-ups by 1 to 2 orders of magnitude \cite{Khanna:2010}, it is not unreasonable to expect similar performance gains for effective source calculations.

There is yet another intriguing prospect for improving calculations involving an effective source. As discussed in Sec.~\ref{sec:no-effsource}, the effective source may be viewed as merely a correction for the fact that the singular field is not known exactly. This begs the question of whether the singular field could be calculated exactly on a world-tube boundary. Not only would this improve convergence in a numerical code (arbitrarily high convergence in grid spacing, exponential convergence in $l$ or $m$ mode sums), but it would also negate the need to calculate an effective source at all. The entire computational cost of implementing the effective source approach would be in the computation of the value of the singular field on the boundary. While an exact calculation of the singular field may not be realistic, one should recall that from a numerical perspective a value which is correct in the first $16$ digits of a double precision number is effectively `exact' in that further refinements do not change the result. Given the availability of high order expansions of the Green function \cite{Anderson:2003,Anderson:2003:err1,Anderson:2003:err2,Anderson:Eftekharzadeh:Hu:2006,QL,Transport,Wardell:PhD} along with the fact that multi-domain spectral methods \cite{Canizares:Sopuerta:2009,Canizares:Sopuerta:Jaramillo:2010} or adaptive mesh refinement \cite{Thornburg:2009,Thornburg:2010} allow the world-tube boundary to be placed very close to the particle, it seems like this may be a plausible approach, although further investigation is required to determine whether this is truly the case.

Yet another potential optimisation arises from the covariant treatment of Sec.~\ref{sec:effsrc-approx}. Near the particle, the covariantly re-expanded effective source is a reasonable approximation to the `correct' effective source. However, given that it only requires a first order coordinate expansion, it is dramatically more efficient to evaluate numerically. Furthermore, as the divergences are cancelled analytically, it effectively avoids any need for concern about delicate numerical cancellations. Lastly, as the singular field is constructed in such a way that it and its derivative (i.e.\ the self-force) evaluated at the particle are insensitive to these covariant re-expansions, it is plausible that using the covariant re-expansion throughout the world-tube may be possible. This may lead to an `incorrect' regularized field away from the particle, but with sufficient care could potentially still give the `correct' value for the field and its derivative at the particle.

This work focused on the case of a scalar charge moving in a background spacetime. Of arguably much more interest are the cases of gravitational or electromagnetic charges. Fortunately, the calculation strategy remains largely unchanged. One can make use of an analogous Detweiler-Whiting gravitational or electromagnetic Green function which has the same Hadamard-type structure. It will still include functions $U(x,x')^{A}{}_{B'}$ and $V(x,x')^{A}{}_{B'}$ which are analogous to their scalar variants and may be calculated in the exact same way \cite{Transport}. Furthermore, the world function, $\sigma$, and its derivatives will remain unchanged from the scalar case. The full details
of this calculation will be developed in a future work.

\section{Acknowledgements}
We are grateful to Sam Dolan and Leor Barack for much helpful interaction and many suggestions during the progress of this work. We also thank Adrian Ottewill, Marc Casals, Jos\'e Luis Jaramillo, Michael Jasiulek and Abraham Harte for insightful discussions.
We thank Eric Ost for valuable assistance with the computer cluster
used for some of the calculations described in this paper.
Finally, we thank participants of the 2010 and 2011 Capra meetings (in Waterloo and Southampton, respectively) -- 
particularly Eric Poisson and Steven Detweiler -- for many illuminating conversations.

\appendix
\section{Covariant expansions} \label{sec:Expansions}
In this appendix, we develop covariant expansion expressions for the biscalars $U(x,x')$, $U(x,x'')$, $\sigma_{a'}(x,x')$, $\sigma_{a''}(x,x')$ and $\int_u^v V(x, z(\tau)) d\tau$  appearing in Eq.~\eqref{eq:PhiS}. We eventually seek expansions about the point $\bar{x}$. In doing so, we follow the strategy of Haas and Poisson \cite{Poisson:2003,Haas:Poisson:2006}:
\begin{itemize}
\item For the generic biscalar $A(x,z(\tau))$, write it as $A(\tau) \equiv A(x,z(\tau))$.
\item Compute the expansion about $\tau = \bar{\tau}$. This takes the form
\begin{equation}
A(\tau) = A(\bar{\tau}) + \dot{A}(\bar{\tau}) (\tau-\bar{\tau}) + \frac{1}{2}\ddot{A}(\bar{\tau}) (\tau-\bar{\tau})^2 + \cdots,
\end{equation}
where $\dot{A}(\bar{\tau}) = A_{;\bar{\alpha}} u^{\bar{\alpha}}$, $\ddot{A}(\bar{\tau}) = A_{;\bar{\alpha} \bar{\beta}} u^{\bar{\alpha}} u^{\bar{\beta}}$, $\cdots$.
\item Compute the covariant expansions of the coefficients $\dot{A}(\bar{\tau})$, $\ddot{A}(\bar{\tau})$, $\cdots$ about $\bar{\tau}$.
\item Evaluate the expansion at the desired point, e.g. $A(x') = A(x,x')$.
\item The resulting expansion depends on $\tau$ through the powers of $\tau-\bar{\tau}$. Replace these by their expansion in $\epsilon$ (about $\bar{x}$), the distance between $x$ and the world-line.
\end{itemize}

A key ingredient of this calculation is the expansion of $\Delta \equiv \tau-\bar{\tau}$ in $\epsilon$. This expansion was developed by Haas and Poisson \cite{Haas:Poisson:2006} to sufficient order for the present calculation for the particular choices $\Delta_+ \equiv v-\bar{\tau}$ and $\Delta_- \equiv u-\bar{\tau}$. They found
\begin{equation}
\label{eq:Delta}
\Delta_\pm = (\rbar \pm \sbar) \mp \frac{(\rbar \pm \sbar)^2}{6\,\sbar} R_{u \sigma u \sigma} \mp \frac{(\rbar \pm \sbar)^2}{24\,\sbar} \left[ (\rbar \pm \sbar)R_{u \sigma u \sigma | u} - R_{u \sigma u \sigma | \sigma}\right] + \mathcal{O}(\epsilon^5).
\end{equation}

\subsection{Expansion of $U(x,x')$ and $U(x,x'')$}
We now compute expansions of $U(x,x')$ and $U(x,x'')$ about $\bar{x}$. Both calculations proceed in the same way and require the expansion of $U(x,\bar{x})$ about $\bar{x}$ which is given by \cite{Ottewill:Wardell:2009}:
\begin{equation}
U(x,\bar{x}) = \Delta^{1/2} (x,\bar{x}) = 1 + \frac{1}{12} R_{\sigma \sigma} - \frac{1}{24} R_{\sigma \sigma | \sigma} + \mathcal{O}(\epsilon^4).
\end{equation}

Writing $U(\tau) \equiv U(x,z(\tau))$, where $\tau$ stands for either $u$ or $v$, we compute its expansion about $\tau = \bar{\tau}$:
\begin{equation}
\label{eq:Utau}
U(\tau) = U(\bar{\tau}) + \dot{U}(\bar{\tau}) (\tau-\bar{\tau}) + \frac{1}{2}\ddot{U}(\bar{\tau}) (\tau-\bar{\tau})^2  + \frac{1}{6}\dddot{U}(\bar{\tau}) (\tau-\bar{\tau})^3 + \mathcal{O}(\epsilon^4),
\end{equation}
where
\begin{IEEEeqnarray}{rCl}
\label{eq:Ubar}
U(\bar{\tau}) &=& 1 + \frac{1}{12} R_{\sigma \sigma} - \frac{1}{24} R_{\sigma \sigma | \sigma} + \mathcal{O}(\epsilon^4) \\
\label{eq:Ubardot}
\dot{U}(\bar{\tau}) = U_{;\bar{\alpha}} u^{\bar{\alpha}} &=& \frac{1}{6} R_{u \sigma} - \frac{1}{12}R_{u \sigma | \sigma} + \frac{1}{24} R_{\sigma \sigma | u}  + \mathcal{O}(\epsilon^3)  \\
\label{eq:Ubarddot}
\ddot{U}(\bar{\tau}) = U_{;\bar{\alpha} \bar{\beta}} u^{\bar{\alpha}} u^{\bar{\beta}} &=& \frac{1}{6} R_{uu} - \frac{1}{12} R_{u u |\sigma} + \frac{1}{6} R_{u\sigma |u} + \mathcal{O}(\epsilon^2)\\
\label{eq:Ubardddot}
\dddot{U}(\bar{\tau}) = U_{;\bar{\alpha} \bar{\beta} \bar{\gamma}} u^{\bar{\alpha}} u^{\bar{\beta}} u^{\bar{\gamma}}&=& \frac{1}{4} R_{uu|u} + \mathcal{O}(\epsilon).
\end{IEEEeqnarray}
Substituting Eqs.~\eqref{eq:Delta} and \eqref{eq:Ubar}-\eqref{eq:Ubardddot} into \eqref{eq:Utau} and evaluating at $\tau = \{u, v\}$, we get our final expression for the expansion of $U_- \equiv U(x,x')$ and $U_+ \equiv U(x,x'')$ about $\bar{x}$:
\begin{IEEEeqnarray}{rCl}
\label{eq:Upm}
U_\pm &&= 1 + \frac{1}{12}\bigg[R_{\sigma \sigma} + 2(\rbar \pm \sbar)R_{u \sigma} + (\rbar \pm \sbar)^2 R_{uu} \bigg] \nonumber \\
&& +\: \frac{1}{24} \bigg[-R_{\sigma \sigma | \sigma} + (R_{\sigma \sigma | u}-2R_{u \sigma | \sigma})(\rbar \pm \sbar) + (2R_{u \sigma | u} -R_{u u | \sigma}) (\rbar \pm \sbar)^2 + R_{u u | u} (\rbar \pm \sbar)^3 \bigg] \nonumber \\
&&+ \mathcal{O}(\epsilon^4).
\end{IEEEeqnarray}
The first term here is $\mathcal{O}(1)$, the second term is $\mathcal{O}(\epsilon^2)$ and the third term is $\mathcal{O}(\epsilon^3)$. Note that for vacuum spacetimes, these become $U(x,x') = 1 + \mathcal{O}(\epsilon^4) = U(x,x'')$, as is to be expected. Additionally, note that the difference between $U(x,x')$ and $U(x,x'')$ first becomes apparent at $\mathcal{O}(\epsilon^2)$.

\subsection{Expansion of $\sigma_{\alpha'}u^{\alpha'}$ and $\sigma_{\alpha''}u^{\alpha''}$ }
Haas and Poisson give expansions for $\sigma_{\alpha'}u^{\alpha'}$ and $\sigma_{\alpha''}u^{\alpha''}$. They are:
\begin{eqnarray}
\label{eq:r_ret}
\sigma_{\alpha'} u^{\alpha'} &=& \sbar - \frac{\rbar^2-\sbar^2}{6\,\sbar} R_{u \sigma u \sigma} - \frac{\rbar-\sbar}{24\,\sbar} \left[ \left(\rbar-\sbar\right)\left(\rbar+2\,\sbar\right) R_{u \sigma u \sigma | u} - \left(\rbar+\sbar\right) R_{u \sigma u \sigma | \sigma} \right] + \mathcal{O}(\epsilon^5)\qquad
\end{eqnarray}
\begin{eqnarray}
\label{eq:r_adv}
- \sigma_{\alpha''} u^{\alpha''} &=& \sbar - \frac{\rbar^2-\sbar^2}{6 \,\sbar} R_{u \sigma u \sigma} - \frac{\rbar+\sbar}{24\,\sbar} \left[ \left(\rbar+\sbar\right)\left(\rbar-2\,\sbar\right) R_{u \sigma u \sigma | u} - \left(\rbar-\sbar\right) R_{u \sigma u \sigma | \sigma} \right] + \mathcal{O}(\epsilon^5),\qquad
\end{eqnarray}
In these expressions, the first term is  $\mathcal{O}(\epsilon)$, the second is $\mathcal{O}(\epsilon^3)$ and the third is $\mathcal{O}(\epsilon^4)$. Note that difference between $\sigma_{\alpha'} u^{\alpha'}$ and $\sigma_{\alpha''} u^{\alpha''}$ only becomes apparent at $\mathcal{O}(\epsilon^4)$.

\subsection{Expansion of $\int_u^v V(x,z(\tau)) d\tau$}
The expansion of the tail term in Eq.~\eqref{eq:PhiS} poses an additional potential difficulty because of the integration over a portion of the world-line. However, expanding $V(x,z(\tau))$ about $\bar{x}$, the integration becomes a trivial integration of powers of $\tau$.

In the following, we make use of the expansion of $V(x,\bar{x})$ about $\bar{x}$,
\begin{equation}
V(x,\bar{x}) = \frac{1}{2} (\xi-\frac{1}{6}) \bar{R} - \frac{1}{4}(\xi-\frac{1}{6}) \bar{R}_{|\sigma} + \mathcal{O}(\epsilon^2).
\end{equation}
We now proceed, as before, by defining $V(\tau) \equiv V(x,z(\tau)$, where $\tau$ lies between $u$ and $v$, and computing the expansion about $\tau = \bar{\tau}$:
\begin{equation}
\label{eq:Vtau}
V(\tau) = V(\bar{\tau}) + \dot{V}(\bar{\tau}) ( \tau - \bar{\tau}),
\end{equation}
where
\begin{IEEEeqnarray}{rCl}
\label{eq:Vbar}
V(\bar{\tau}) &=& \frac{1}{2} (\xi-\frac{1}{6}) \bar{R} - \frac{1}{4}(\xi-\frac{1}{6}) \bar{R}_{|\sigma} + \mathcal{O}(\epsilon^2) \\
\label{eq:Vbardot}
\dot{V}(\bar{\tau}) = V_{;\alpha} u^\alpha &=& \frac{1}{4}(\xi-\frac{1}{6}) \bar{R}_{|u} + \mathcal{O}(\epsilon).
\end{IEEEeqnarray}
The integration along the world-line is now straightforward since the only dependence of the integrand on $\tau$ comes through the factor $\tau - \bar{\tau}$. Performing the integration and substituting Eqs.~\eqref{eq:Delta}, \eqref{eq:Vbar} and \eqref{eq:Vbardot} into the result, we get our final expression for the expansion of $\int_u^v V(x,z(\tau)) d\tau$ about $\bar{x}$:
\begin{IEEEeqnarray}{rCl}
\label{eq:intV}
\int_u^v V(x,z(\tau)) d\tau \approx (\xi-\frac{1}{6})\bar{R} \,\sbar + \frac{1}{2}(\xi-\frac{1}{6})(\bar{R}_{|u}\,\rbar\,\sbar-\bar{R}_{|\sigma}\,\sbar) + \mathcal{O}(\epsilon^3).
\end{IEEEeqnarray}
The first term here is $\mathcal{O}(\epsilon)$ and the second term is $\mathcal{O}(\epsilon^2)$. Note that for vacuum spacetimes, $V(x,z(\tau)) = \mathcal{O}(\epsilon^4)$ and this term does not contribute to the singular field until $\mathcal{O}(\epsilon^5)$.

\section{Leading-order piece of the coordinate expression for $\sbar^2$}
\label{sec:s2positive}

As we discuss in Sec. \ref{sec:PhiS-practical}, it is highly desirable for a coordinate 
representation of the effective source to not diverge anywhere. Unfortunately, a common feature of series expansions is that they have a finite region of validity (for Taylor series, this is denoted
by their radius of convergence). Outside this region, spurious singularities tend to appear.
As already indicated, a re-expansion of the denominator of the singular field (leaving only its quadratic leading-order dependence on the coordinate separation and bringing 
all higher-order terms up to the numerator) allows one to avoid most potential singularities away from the 
position of the particle. Here we demonstrate this explicitly for the case of Schwarzschild coordinates.

In these coordinates, the leading-order dependence of the denominator (essentially given by 
$\sbar^2 \equiv (g^{\bar{\alpha}\bar{\beta}}+u^{\bar{\alpha}}u^{\bar{\beta}})\sigma_{\bar{\alpha}}\sigma_{\bar{\beta}}$) on the coordinate separations is  
\begin{align}
\sbar^2 = &-F(1-F(u^t)^2)(\Delta t)^2 - 2u^tu^r(\Delta t)(\Delta r) - 2FR^2u^tu^\phi (\Delta t)(\Delta \phi) 
 \nonumber \\ &+R^2 (\Delta \theta)^2 + \frac{F+(u^r)^2}{F^2}(\Delta r)^2 + \frac{2R^2}{F}u^r u^\phi (\Delta r)(\Delta \phi)  
+ R^2(1+R^2(u^\phi)^2)(\Delta \phi)^2,
\end{align}
where $\Delta t = t - \bar{t}, \Delta r = r - R,$ etc. (recalling that barred coordinates 
refer to the position of the particle), $R$ is the radial position of the particle in Schwarzschild coordinates, and $F := 1-2M/R$.

If we take $\Delta t = 0$, this reduces to 
\begin{align}
\sbar^2(\Delta t = 0) = R^2 (\Delta \theta)^2 + \left[ \frac{F+(u^r)^2}{F^2}(\Delta r)^2 + \frac{2R^2}{F}u^r u^\phi (\Delta r)(\Delta \phi)  
+ R^2(1+R^2(u^\phi)^2)(\Delta \phi)^2 \right].
\end{align}
The condition $\Delta t = 0$ imposes that the position and four-velocity of 
the particle are evaluated at the same coordinate time as where the effective source is evaluated, 
or in other words, the particle location and field point need to be at the same $t$-hypersurface.

All except the cross term $\propto (\Delta r) (\Delta \phi)$ are manifestly positive-definite. The combination of 
terms in the square brackets, however, can also be shown to be positive-definite; it is a quadratic form
in $\{\Delta r, \Delta \phi\}$: 
\begin{equation}
    A(\Delta r)^2 + B(\Delta r)(\Delta \phi) + C(\Delta \phi)^2 
\end{equation}
where 
\begin{equation}
    A =  \frac{(F+(u^r)^2)}{F^2},\,\,\, B =  \frac{2R^2}{F}u^r u^\phi,\,\,\, C= R^2(1+R^2(u^\phi)^2)
\end{equation}
This will be positive-definite if $B^2 -4AC < 0$. The condition is easily verified to reduce to
\begin{equation}
    0 < F + (u^r)^2 + FR^2(u^\phi)^2,
\end{equation}
which is true for $F>0$. Thus, $s^2(\Delta t =0)$ is positive everywhere except that it vanishes at the 
position of the particle. This implies that the re-expanded singular field, which keeps only 
the quadratic dependence on the coordinate separation in its denominator, diverges only at the 
location of the particle, and consequently, that the corresponding effective source is regular everywhere else.

\section{Efficient numerical computation of the singular field and effective source}
\label{sec:numerical}
The calculation of numerical values for $\tilde{\Phi}_{\rm S}$ and $S_{\rm eff}$ requires the numerical evaluation of their coordinate expansion. This amounts to numerically evaluating a multivariate polynomial (in $(\Delta r, \Delta \theta, \Delta \phi)$) with coefficients which are potentially complicated functions of the particle's location and four-velocity. Furthermore, in a numerical code this must be done at every point on a $3$D grid!\footnote{Even in $1+1$D and $2+1$D codes this is necessary because of the numerical integration involved.} Clearly, it is crucial to make this evaluation as efficient as possible, so that the computational cost of the effective source does not prohibit its use in a numerical code.

Fortunately, there are a two points which enable significant improvements:
\begin{itemize}
\item Since the expansions are all about $\bar{x}$, the coefficients of the polynomial do not change from grid point to grid point. They may change from one iteration to the next, however.
\item In some cases such as with circular orbits in Schwarzschild and Kerr spacetimes, this change between iterations is trivial and does not necessarily require recalculation of the effective source.
\end{itemize}
This suggests an obvious optimization. The coefficients are only computed once at the start of an iteration and then their numerical values are stored. The evaluation at each grid point then becomes simple multiplication by powers of $(\Delta r, \Delta \theta, \Delta \phi)$; a relatively fast and computationally efficient operation. A further optimisation can be found by computing powers of $(\Delta r, \Delta \theta, \Delta \phi)$ only once at the start of the simulation, providing the grid structure does not change. Altogether, this yields an enormous speed improvement - a factor of $50-100$ in many cases. Similar tricks may also be employed with other parameters (mass, spin, etc.) which do not change through the lifetime of the simulation. Furthermore, if accuracy is important and delicate numerical cancellations are causing problems, this approach allows for the use of highly accurate methods such as Kahan \cite{Kahan:1965} summation to minimize problems arising from numerical round-off.

In addition to the numerical algorithm, it also important to consider the method for generating the code. Given the length of the expressions, it is impractical to manually type them in. Instead, we have directly generated \emph{C} code from the \emph{Mathematica} expressions and have made both available online \cite{EffSource-online}.

\bibliography{EffSource}{}

\end{document}